\documentclass[aps,prx,reprint,superscriptaddress,longbibliography]{revtex4-2}

\usepackage[dvipsnames]{xcolor}

\usepackage{amsmath,amsthm, bm,bbm, graphicx}
\usepackage{amssymb, amsfonts}
\usepackage{lineno}

\newcommand{\J}{\mathbf{J}}

\newcommand{\Jm}{\bm{J}}

\newcommand{\eye}{\bm{I}}
\newcommand{\ob}{f}
\newcommand{\symfactor}{\nu}
\newcommand{\symsymbol}{s}

\begin{document}

\title{Solution of a chaotic neural network at fixed connectivity}
\author{Albert J. Wakhloo}
\affiliation{Center for Theoretical Neuroscience, Zuckerman Institute, Columbia University}
\email{ajw2232@cumc.columbia.edu}
\begin{abstract}
We calculate the moments and response functions of a nonlinear random recurrent neural network in the large-$N$ limit using a diagrammatic technique. Our approach does not require averaging over synaptic weights and gives the first nontrivial term in a $1/\sqrt{N}$ expansion of general intensive-order correlation functions, proving a recent conjecture by Shen and Hu as a special case. Our results provide an analytical link between synaptic connectivity, correlations in spontaneous activity, and the response of a network to small perturbations.
\end{abstract}

\maketitle

\section{Introduction}

Random recurrent neural networks are one of the only analytically tractable models of strongly recurrent dynamics, and they provide a natural setting to study how synaptic connectivity and external inputs shape neural activity \cite{sompolinsky1988chaos, rajan2010stimulus, KadmonSompolinsky2015, mastrogiuseppe2018linking}. 
In this model, a collection of $N$ neurons communicate with one another through a set of randomly chosen connections. The preactivations of the units are denoted by $ x_i (t)
$ and the random connections by a matrix $J_{ij}$ with $1\leq i,j\leq N$. The dynamics follow
\begin{gather}
    \partial_t x_i = -x_i + \sum_j J_{ij} \phi(x_j) + \xi_i,
    \label{eq:system}
\end{gather}
where $ \xi_i(t)$ are external inputs into the neurons, and $\phi(\cdot)$ is a transfer function mapping the $x_i$ to normalized firing rates. If the random connections $J_{ij}$ have a sufficiently large variance and the inputs do not overwhelm the influence of the recurrence, the activity of such networks is generically chaotic \cite{sompolinsky1988chaos, rajan2010stimulus}. 

The dynamics of random recurrent networks have been extensively studied using dynamical mean-field theory \cite{sompolinsky1988chaos, rajan2010stimulus, KadmonSompolinsky2015, mastrogiuseppe2018linking, clark2023dimension, crisanti2018path}. As described in Refs.\ \cite{sompolinsky1988chaos, crisanti2018path, helias2020statistical}, if the $J_{ij}$ are drawn independently from a zero-mean distribution with variance $\langle J_{ij}^2 \rangle_{\bm J}  = \sigma^2/N$ and the $\xi_i$ are stationary i.i.d.\ Gaussian processes, then to zeroth order in $ 1/\sqrt{N} $, the neurons in the network decouple. The statistics of any individual neuron are identical to those of the scalar system:
\begin{gather}
    \partial_t x = -x + \eta + \xi,
    \label{eq:mft-one}
\end{gather}
where $\xi$ is a draw from the same Gaussian process as the $\xi_i$, and $ \eta $ is a zero-mean Gaussian process with autocovariance $\langle \eta(t) \eta(t') \rangle_\eta = \Delta(t-t')$. The autocovariance $\Delta$ is fixed by the self-consistency condition
\begin{gather}
    \Delta(t-t') = \sigma^2 \langle \phi(x(t)) \phi(x(t')) \rangle_{\xi, \eta}.
    \label{eq:dmft-selfcons} 
\end{gather}
Using the mean-field equations \eqref{eq:mft-one} and \eqref{eq:dmft-selfcons}, it is possible to calculate various self-averaging statistics of the network activity---i.e., statistics that do not depend on the specific realization of the synaptic connections $J_{ij}$. 

A limitation of traditional dynamical mean-field approaches is that they do not provide statistics that depend on the \emph{specific} choice of $J_{ij}$, such as the pairwise correlations of the units, or the response of a unit $x_i$ to a perturbation of another unit  $x_j$. In other words, it is not possible to calculate quantities that are not self-averaging. In this work, we address this gap by proving formulae for general correlation-response functions of the network activity at a fixed instance of the synaptic connections.

A key motivation for our work is a recent conjecture from Shen and Hu. In Ref.\ \cite{shen2025covariance}, the authors proposed that the pairwise correlations between the units of the nonlinear network Eq.\ \eqref{eq:system} obey simple equations at a fixed realization of the connectivity $J_{ij}$---see Eq. $\eqref{eq:introshenhu}$ below. The authors provided numerical evidence for these equations, but they did not prove them. These formulae suggest that it might be possible to analyze random recurrent networks at a fixed instance of the synaptic connections. Here we show that this is indeed the case: the Shen-Hu formulae are special cases of a more general fixed-connectivity result.

The proof of this result relies on a diagrammatic expansion. As described below, we Taylor expand a path integral representation of the nonlinear network's dynamics around the mean-field solution. We then use diagrams to identify the leading terms in the expansion of arbitrary $m$-point correlation and response functions. Resumming the remaining series establishes the result.

This paper is organized as follows. We first summarize our results in Sec.\ \ref{sec:summary}. In Sec.\ \ref{sec:setup}, we introduce the Martin-Siggia-Rose-De Dominicis-Janssen (MSRDJ) path integral formulation of the network dynamics, and we define the diagrammatic formalism used in the proof. We present several example diagrams that illustrate the idea of the calculation that follows. Finally, Sec.\ \ref{sec:proof-sketch} contains a sketch of the main steps of the proof, and Sec.\ \ref{sec:proof} presents the proof itself. 

\section{Summary of results}
\label{sec:summary}

In this work, we prove a Wick formula for general moments and response functions of the network Eq.\ \eqref{eq:system}, without averaging over the connections $J_{ij}$. While our formula implies the Shen-Hu conjecture Eq. \eqref{eq:introshenhu} as a special case, our main result is an equation for general correlation functions of the form 
\begin{gather} 
\begin{aligned}
\bigg\langle 
\frac{\delta^q}
{\delta \xi_{b_1}(t'_1) \cdots \delta \xi_{b_q} (t'_q)}
g_1(x_{a_1}(t_1)) \cdots g_p (x_{a_p} (t_p)) 
\bigg\rangle,
\end{aligned}
\label{eq:corrfunc}
\end{gather} 
for arbitrary fixed $q,p$. Here, the $g_\mu$ are arbitrary odd functions, the average is over $\xi_i$, and the indices $1\leq a_\mu, b_\nu \leq N$ can take arbitrary values. Statistics for the undriven network follow by taking $\langle \xi_i(t)\xi_i(t')\rangle \to 0^+$. In what follows, we assume that the nonlinearity $\phi(\cdot)$ is odd, and we work in the large-$N$ regime, $N \gg 1$. 

While we derive the formula for Eq.\ \eqref{eq:corrfunc} in the general case, we note that the moments of the pre- or post-activations, $\langle x_{a_1} \cdots x_{a_p}\rangle, \langle \phi(x_{a_1}) \cdots \phi(x_{a_p})\rangle,$ follow by setting the functions $g_\mu$ to be one of $g_\mu(x) =x, \phi(x)$ and setting $q=0$ above. Our results therefore describe how synaptic weights shape the moments and responses of a random recurrent neural network at a fixed instance of the synaptic connectivity.

We first consider the two-point functions. In Sec.\ \ref{sec:resum}, we present a general formula for correlation functions of the form Eq.\ \eqref{eq:corrfunc} with $p+q=2$ (see Eqs.\ \eqref{eq:ondiag} and \eqref{eq:offdiagtwopoint}). The formula for general odd $g_\mu $ is somewhat complicated, so we focus here on the two-point functions for the pre- and post-activations, $x, \phi(x)$. In this special case, we show that the general formula implies the conjecture of Shen and Hu \cite{shen2025covariance}. 

More precisely, let us define the pairwise correlation functions $C^x_{ij}(t-t') = \langle x_i(t) x_j(t')\rangle_{\xi}$ and similarly for $C^\phi_{ij}(t-t')$. In Ref.\ \cite{shen2025covariance}, the authors conjectured that these functions are given by the inverse Fourier transform of
\begin{align}
    \bm C^x(\omega) 
    &= h_x(\omega)\bm J (\bm I - R^\phi_0 \bm J)^{-1}
    (\bm I - R^\phi_0 \bm J)^{-\dagger} \bm J^\top  
    \label{eq:introshenhu}
    \\ 
    &\quad + \frac{G(\omega)}{1+\omega^2}  
    (\bm I - R^\phi_0 \bm J)^{-1}
    (\bm I - R^\phi_0 \bm J)^{-\dagger}, 
    \nonumber 
    \\
    \bm C^\phi(\omega) 
    &= h_\phi(\omega)(\bm I - R^\phi_0 \bm J)^{-1} 
    (\bm I - R^\phi_0 \bm J)^{-\dagger},
    \nonumber 
\end{align}
where the scalar prefactors are
\begin{align}
    h_x(\omega) &= \frac{C^\phi_0(\omega) - \langle \phi'\rangle_0^2 C^x_0(\omega)}{1+\omega^2},
    \\
    h_\phi(\omega) &= C^\phi_0(\omega) (1-\sigma^2 |R^\phi_0|^2).\nonumber
\end{align}
Here, $\bm C^x(\omega), \bm C^\phi(\omega)$ are the Fourier transforms of the pairwise correlation functions, $C_{ij}^\phi (\omega) = \int d\tau e^{-i\omega \tau} \langle \phi(x_i(t+\tau)) \phi(x_j(t))\rangle_{t, \xi}$ and similarly for $C^x_{ij}(\omega)$. In the above, we use $\bm I$ to denote the identity matrix, and the function $G(\omega)$ is the Fourier transform of the noise autocovariance: $G(\omega) = \int d\tau e^{-i\omega \tau} \langle \xi_i(t+\tau) \xi_i(t)\rangle_{t,\xi}$. The scalar-valued functions $C^\phi_0, C^x_0,$ and $R^\phi_0$ are statistics of the single-site mean-field theory. The functions $C^\phi_0(\omega), C^x_0(\omega) $ are the Fourier transformed autocovariance functions of $\phi(x)$ and $x$, respectively, $\langle\phi'\rangle_0 = \langle  \phi'(x)\rangle_{\xi, \eta}$ is the gain, and $R^\phi_0 = \langle \phi' \rangle_0/(1+i\omega)$ is the post-activation linear response, with the averages over $x$ taken with respect to the stochastic process in Eq.\ \eqref{eq:mft-one}. 

Our results show that the equations \eqref{eq:introshenhu} are valid to leading non-trivial order in $1/\sqrt{N}.$ That is,  the error in the equations \eqref{eq:introshenhu} is $O(N^{-1/2})$ on the diagonal and $O(N^{-1})$ off it.

The same general formula for the two-point functions implies simple equations for the linear response functions, defined as $R^\phi_{ij}(t-t') = \langle \delta \phi(x_i(t)) /\delta \xi_{j}(t')\rangle $ and similarly for $R^x_{ij}$. These functions describe the response of a neuron $i$ to a perturbation of neuron $j$.  We show that in Fourier space, the response functions of the pre- and post-activations are
\begin{align}
    \bm R^\phi(\omega) 
    &= R^\phi_0 (\bm I - R^\phi_0 \bm J)^{-1} 
    + O(N^{-1/2} \bm I + N^{-1} \bm 1), 
    \label{eq:responseintro}
    \\ 
    \bm R^x(\omega) 
    &= R^x_0 (\bm I - R^\phi_0 \bm J)^{-1}  
    + O(N^{-1/2} \bm I + N^{-1} \bm 1), 
    \nonumber 
\end{align}
where $\bm 1$ is the matrix of all ones, $R^x_0 = 1/(1+i\omega)$ is the preactivation linear response under Eq.\ \eqref{eq:mft-one}, and the error estimates should be understood elementwise. That is, the error is again $O(N^{-1/2})$ for the diagonal entries and $O(N^{-1})$ for the off-diagonals. Neglecting $O(N^{-1/2})$ fluctuations, the diagonal entries of Eqs.\ \eqref{eq:introshenhu} and  \eqref{eq:responseintro} reduce to the single-site mean-field equations. Off the diagonal, they express neural correlations and response properties in terms of the connections $J_{ij}$. These equations therefore describe how synaptic connections shape pairwise correlations and linear responses in the network.

We now state the result for higher-order correlation functions. It is convenient to use a single symbol to represent both the functions $g_\mu(x_{a_\mu}(t_\mu))$ and the derivatives in the average Eq.\ \eqref{eq:corrfunc}. To this end, we define the symbols $f_\mu(t_\mu)$ with $1 \leq \mu \leq p+q$ to be one of these two quantities. That is, $f_\mu$ is equal to either $ g_{\mu-q}(x_{a_{\mu-q}}(t_{\mu-q}))$ or $\delta /\delta \xi_{b_\mu}(t_\mu)$, so that Eq.\ \eqref{eq:corrfunc} can be written as
 \begin{gather}
     \langle f_1(t_1) \cdots f_m(t_m) \rangle ,
 \end{gather}
where $m=p+q$. For odd $m$, the correlation function vanishes, so we take $m$ even. To begin, consider the special case of all $a_\mu, b_\nu$ distinct. In this setting, we establish the Wick formula: 
\begin{align} 
\langle f_1(t_1) \cdots f_m(t_m)\rangle
&= \sum_{p \in P_2(1, \cdots, m)} 
\prod_{(\mu,\nu) \in p} 
\langle f_\mu(t_\mu)f_\nu(t_\nu)\rangle 
\nonumber \\
&\quad + O(N^{-m/4 - 1/2 }),
\label{eq:simplemoment}
\end{align} 
where the two-point functions follow from Eqs.\ \eqref{eq:introshenhu} and \eqref{eq:responseintro} when the nonlinearities $g_\mu$ are one of $x, \phi(x)$ and from Eq.\ \eqref{eq:offdiagtwopoint} more generally. The variable $P_2(1, \cdots, m)$ denotes the set of all partitions of the set $\{1, \cdots , m \}$ into pairs. 

Note that $\langle f_\mu f_\nu\rangle  = O(N^{-1/2})$ off the diagonal, $a_\mu \neq a_\nu$. Thus, the product over Wick pairings scales as $O(N^{-m/4})$. Combining the above with Eqs.\ \eqref{eq:introshenhu} and \eqref{eq:responseintro} therefore gives a closed-form solution to the moments and response functions of $x_i$ and $\phi(x_i)$ with all indices distinct to leading non-trivial order in $1/\sqrt{N}$. 

The case with repeated $a_\mu$ indices requires slightly more care, essentially because averages of the form $\langle g_1(y) \cdots g_l(y)\rangle_y$ for Gaussian $y$ do not factorize into products of two-point functions. In this case, we prove in Sec.\ \ref{sec:resum} that a similar Wick formula holds, with the provision that same-site averages are not factorized into products of two-point functions---see Eq.\ \eqref{eq:mpoint} for the full formula. This gives a closed-form solution to intensive-order correlation functions for general odd $g_\mu$. Our results therefore link connectivity, correlations, and response functions in a nonlinear random recurrent neural network.

\section{Setup} 
\label{sec:setup}
\subsection{Generating functional}
In what follows, we will assume that $\phi(\cdot)$ is an odd function and that the connections $J_{ij}$ are drawn i.i.d.\ from a zero-mean probability distribution, $\langle J_{ij}\rangle_{\bm J}= 0$ with $\langle J_{ij}^2\rangle_{\bm J} = \sigma^2/N$. We assume that the connections scale as $N^{-1/2}$ so that for all integers $k$: 
\begin{gather}
    \lim_{N\to\infty}\langle |\sqrt{N}J_{ij}|^k \rangle_{\bm J} < \infty.
    \label{eq:moment}
\end{gather}
We do not require that the $J_{ij}$ are Gaussian. We additionally assume that the driving fields $\xi_i$ are independent draws from a zero-mean Gaussian process with autocovariance $\langle \xi_i(t) \xi_i(t')\rangle_\xi = G(t-t').$

Our proof relies on the MSRDJ generating functional under the Ito convention \cite{martin1973statistical, hertz2016path}. In this formalism, the expectation of a correlation function of the form in Eq.\ \eqref{eq:corrfunc} is represented as 
\begin{widetext}
\begin{align}
    \langle f_1 (t_1) \cdots f_m(t_m)\rangle_{\xi} 
    &= \bigg\langle 
    \int\bigg(\prod_i  D x_i  \delta([1+\partial_t] x_i - \sum_j J_{ij} \phi(x_j) - \xi_i)\bigg) 
    f_1 (t_1) \cdots f_m(t_m) 
    \bigg\rangle_\xi.
\end{align}
\end{widetext}
Expressing the delta functions as a Fourier transform and performing the Gaussian integral over the $\xi$ variables gives 
\begin{gather}
    \int\bigg(\prod_i  D x_i D \hat x_i\bigg) e^{S[\bm x,\bm{\hat x}]} f_1 (t_1) \cdots f_m(t_m) , 
\end{gather}
where the MSRDJ action is 
\begin{align}
    S[\bm x,\bm {\hat x} ] 
    &= \sum_{i} \bigg [
    \int dt  -i\hat x_i(t) (1+\partial_t) x_i(t) 
    \nonumber \\
    &\qquad
    + \sum_j i \hat x_i(t) J_{ij} \phi_j(t) 
    \nonumber \\
    &\qquad
    - \frac{1} 2 \int dt dt'  
    \hat x_i(t) G(t-t') \hat x_i(t') 
    \bigg],
    \nonumber
\end{align}
and we introduced the shorthand $\phi_j = \phi(x_j).$ A useful feature of this formalism is that correlations involving $i\hat x_i$ generate response functions; see e.g.\ Ref.\ \cite{hertz2016path}. For example, 
\begin{gather}
    R^x_{ij}(\tau) = \bigg\langle \frac{\delta x_i(t+\tau)}{\delta \xi_j (t)}\bigg\rangle_{t,\xi} = \langle i \hat x_j(t) x_i(t+\tau)\rangle_{t, \xi},
\end{gather}
where $\bm R^x(\tau)$ is the linear response function. In general, each factor of $i\hat x_j$ in a correlation function brings a factor of $\delta / \delta \xi_j$ to the average. Correlation functions of the form Eq.\ \eqref{eq:corrfunc} can therefore be represented as moments of $g_\mu(x_{a_\mu})$ and $i\hat x_{b_\nu}$.

We  expand the generating functional around the mean-field solution.  To motivate this, note that the dynamical mean-field solution captures the self-averaging statistics of the network up to $O(1)$. Given that we are interested in the large-$N$ behavior of the network, the dynamical mean-field solution provides a natural background to expand against. To do this, note that the action associated with $N$ independent copies of the scalar system Eq.\ \eqref{eq:mft-one} is given by: 
\begin{gather}
    S_0[\bm x,\bm{\hat x}] = \sum_i -i\hat x_i (1+\partial_t) x_i - \frac 1 2 \hat x_i (\Delta+G) \hat x_i, 
\end{gather}
where above and in what follows we suppress the explicit time integrals. Adding and subtracting $-\frac 12 \hat x_i \Delta \hat x_i$, we can write the full nonlinear action $S$ as a sum of free and interacting parts
\begin{align}
    S[\bm x, \bm{\hat x}] 
    &= \sum_i 
     \underbrace{-i\hat x_i (1+\partial_t) x_i - \frac 1 2 \hat x_i (\Delta+G) \hat x_i}_{S_0}
     \nonumber \\
     &\qquad
     + \underbrace{\frac 12 \hat x_i \Delta \hat x_i + \sum_j i\hat x_i J_{ij}\phi_j }_{S_{int}}.
\end{align}
Expanding $S_{int}$, we write the correlation function $\langle f_1 \cdots f_m \rangle$ as an average with respect to the mean-field action, $S_0$: 
\begin{align}
    \langle  f_1 (t_1) \cdots f_m(t_m) \rangle  
    &= \sum_{k,l \geq 0} \frac{1}{k! l!} 
    \bigg\langle  f_1 (t_1) \cdots f_m(t_m) 
    \nonumber \\
    &\;
    \bigg(\sum_{i,j=1}^N i\hat x_i J_{ij}\phi_j\bigg)^k\!
    \bigg(\frac 1 2 \sum_{i=1}^N \hat x_i \Delta \hat x_i\bigg)^l 
    \bigg\rangle_0,
    \label{eq:series}
\end{align}
where $\langle \cdot \rangle$ denotes the average under the action $S$ and $ \langle \cdot \rangle_0 $ the average under $S_0$.

\subsection{Diagram definitions}
\label{sec:defns}

We use diagrams to identify the leading order terms in the series expansion above. As described below, the large-$N$ scaling of the various terms in Eq.\ \eqref{eq:series} depends on the pattern of repeated indices. We use diagrams to group the terms in the series based on this pattern, and we show that their large-$N$ behavior is encoded in the corresponding diagram's structure. In this way, the use of diagrams converts the problem of identifying the leading order terms in a complicated series expansion into a tractable combinatorics problem. This allows us to analyze the large-$N$ behavior of Eq.\ \eqref{eq:system} without first averaging over $J_{ij}$. 

In what follows, we consider the series expansion of an arbitrary $m$-point correlation function, $ \langle f_1(x_{a_1}, \hat x_{a_1}) \cdots f_m(x_{a_m}, \hat x_{a_m})\rangle$ for $m$ even. We set each $f_\mu$ to be one of $f_\mu(x, \hat x) = g_\mu(x)$ with $g_\mu$ odd or $ f_\mu(x,\hat x) =   i\hat x $. The $a_\mu$ indices are held fixed and can take any value from $1$ to $N$ so that repeated indices $a_\mu=a_\nu$ are allowed. We will assume that there are $r$ unique $a_\mu$ indices.  

We expand a term of order $k,l$ in the series Eq.\ \eqref{eq:series} as 
\begin{gather}
\begin{aligned}
    &\frac{i^k}{2^l k! l!} \sum_{\mathbf{i}}\sum_{\mathbf{j}} 
   \langle f_{1} \cdots f_{m}  
   \hat x_{i_1} J_{i_1 i_2} \phi_{i_2}\cdots 
    \hat x_{i_{2k-1}} J_{i_{2k-1} i_{2k}} \phi_{i_{2k}}  
    \\
    &\qquad
    \hat x_{j_1} \Delta \hat x_{j_1}\cdots 
    \hat x_{j_l} \Delta \hat x_{j_l}
    \rangle_0 ,
\end{aligned}
    \label{eq:exampleord}
\end{gather}
where $\mathbf{i} = (i_1, i_2), \cdots,(i_{2k-1},  i_{2k}) $ and $\mathbf{j} = j_1, \cdots, j_l$ are vectors of indices. 

\begin{figure}[t]
    \centering    \includegraphics[width=0.6\linewidth]{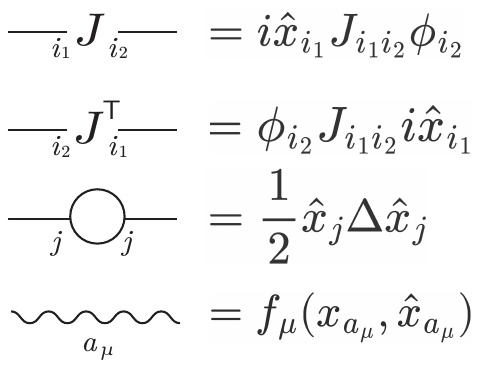}
    \caption{Diagram definitions.}
    \label{fig:defn}
\end{figure}

We use diagrams to represent sums over subsets of index vectors $\mathbf{i}, \mathbf j$ based on the pattern of repeated indices. At a given order $k,l$ of the expansion Eq.\ \eqref{eq:series}, one draws $k$ ``$J_{ij}$ gaps," $l$ ``$\Delta$ bubbles," and $m$ wavy lines, labeling the vertices with indices. Each diagram then represents the sum over all $\mathbf i, \mathbf j$ such that indices connected by an edge in the graph are equal to each other. Stacking the indices into a single vector, $\mathbf{p} = (\mathbf i, \mathbf j)$, a given diagram $D$ represents the sum over the set $\mathbf p \in \mathcal A$, where
\begin{gather}
\begin{aligned}
    \mathcal A 
    &= \{\mathbf p  :  p_\alpha = p_\beta 
    \mathrm{\ for \ } p_\alpha, p_\beta 
    \mathrm{\ connected \ by  \ edges.} 
    \\
    &\qquad
    \mathrm{\ Otherwise \ } p_\alpha \neq p_\beta \}.
\end{aligned}
\end{gather}
Since the value of the resulting sum is invariant under any permutation $(i_\alpha, i_{\alpha+1})\leftrightarrow (i_\beta, i_{\beta+1})$ of the $\mathbf i$ index pairs and any permutation $j_\alpha \leftrightarrow j_\beta$ of the $\mathbf j$ indices---i.e. one can choose which $J_{i_\alpha, i_{\alpha+1}}$ of Eq. \eqref{eq:exampleord} corresponds to each gap in the diagram---we will include this as an additional symmetry factor.

Each diagram $D$ corresponds to a sum of the form
\begin{gather}
    D =  \sum_{(\mathbf i, \mathbf j) \in \mathcal A} \frac{i^k \symfactor}{2^l k! l!} 
   \langle f_1 \cdots f_m \hat x_{i_1} J_{i_1 i_2} \phi_{i_2}\cdots 
    \hat x_{j_l} \Delta \hat x_{j_l}
    \rangle_0, 
    \label{eq:D-defn}
\end{gather}
where $\symfactor$ counts the number of unique permutations of $(i_\alpha, i_{\alpha+1})$ index pairs and $j$ indices. To ensure a one to one mapping between sums and diagrams, we take the following conventions. First, we consider a single labeling of each diagram with $(i_\alpha, i_{\alpha+1})$ and $j$ indices. We then treat permutations which result in the same pattern of matched indices $\mathcal A$ as equivalent, and we include only inequivalent permutations in the definition of $\symfactor$---see below for examples. Similarly, we consider any diagram corresponding to the same set $\mathcal A$ equivalent, and we sum over all inequivalent diagrams (i.e. over equivalence classes of diagrams).

Each diagram may be drawn as a sequence of blocks, such that within each block, adjacent $J_{ij}$ gaps share an index. We prove this in the Appendix Sec.\ \ref{sec:ordlemma} and summarize the argument here. We first show that the sum over diagrams in which a $J_{ij}$ gap or a $\Delta$ bubble is not connected by a sequence of edges to a wavy line is zero. To illustrate the argument, note that the MSRDJ functional satisfies 
\begin{gather}
    1 = \int D\bm x D\bm{\hat x} e^{S_0[\bm x, \bm {\hat x}] + \lambda S_{int}[\bm x, \bm {\hat x}]} 
\end{gather}
for all $\lambda\in \mathbb R $ \cite{hertz2016path}. This follows after a Hubbard-Stratonovich transform on the quadratic $\hat x$ term. Differentiating both sides repeatedly with respect to $\lambda$ and setting $\lambda = 0$ generates diagrams without connections to wavy lines and shows that they sum to zero. A similar argument shows that the sum over diagrams containing $J_{ij}$ gaps or $\Delta$ bubbles that are not connected by a sequence of edges to a wavy line is zero (Sec.\ \ref{sec:ordlemma}).  

The next step is to note that all moments $\langle x_i^m \hat x_i^n g(x_i)^p\rangle_0$  with $m+n+p$ odd and $g$ an odd function are zero. This follows from the symmetry of the action $S_0[x,\hat x]$ under $(x, \hat x) \mapsto (-x, -\hat x)$. The vanishing of odd moments implies that each leg of a $J_{ij}$ gap or $\Delta$ bubble (Fig.\ \ref{fig:defn}) must connect to a partner. We additionally note that averages involving only $\hat x$ are zero (see Ref.\ \cite{hertz2016path}): 
\begin{gather}
    \langle i\hat x(t_1) \cdots i\hat x(t_n)\rangle_0 = \bigg\langle \frac{\delta^n}{\delta \xi(t_1) \cdots \delta \xi(t_n)} 1 \bigg\rangle_0 = 0, 
    \label{eq:xhatvanish}
\end{gather}
for any value of $n$. 

Using these constraints, we show that every non-zero diagram may be drawn as a set of blocks $H^1, \cdots, H^{m/2}$ such that within each block, neighboring $J_{ij}$ gaps have connected legs, and wavy lines flank each block: 

\begin{center} 
    \includegraphics[width=\linewidth]{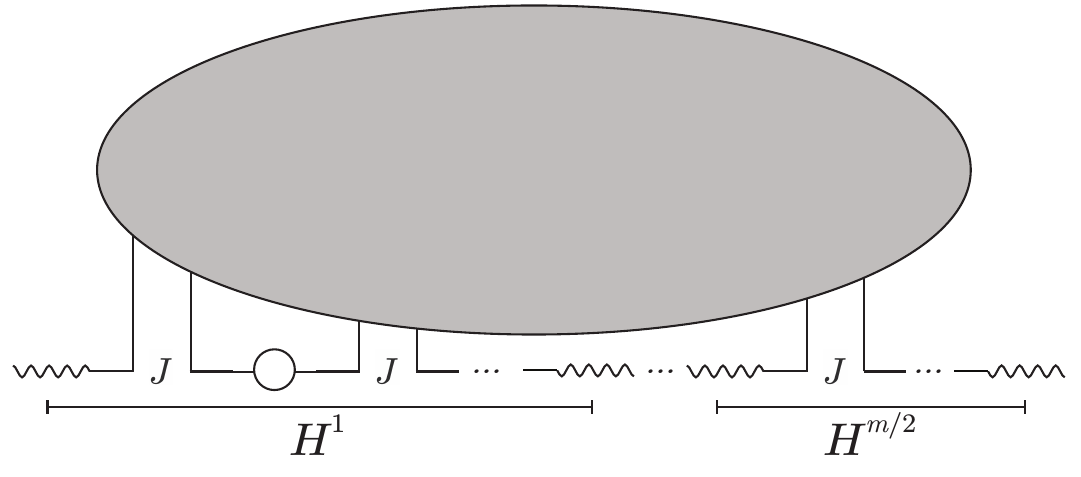}
\end{center}

\noindent where the grey region represents arbitrary connections between the vertices, and there may be an arbitrary number of gaps and bubbles within each block. The wavy lines may also be joined together when the fixed indices repeat, $a_\mu = a_\nu= \cdots = a_\gamma.$

It will be useful to frame the above statement algebraically. In terms of the underlying indices, we show that the $(i_1, i_2), \cdots , (i_{2k-1}, i_{2k})$ can be partitioned into a set of blocks $H^\alpha$ where $1 \leq \alpha \leq m/2$ such that neighboring elements of each block $H^\alpha_n, H^\alpha_{n+1}$ share an index. Formally, each element $H^\alpha_n \in H^\alpha$ is given by
\begin{gather}
    H^\alpha_n = (i_\beta, i_{\beta+1}) \mathrm{ \ or \ } (i_{\beta+1}, i_\beta) , \quad  \quad n = 1,\cdots, q_\alpha,
    \label{eq:hord}
\end{gather}
where $q_\alpha$ denotes the number of $J_{ij}$ gaps in the block so that $\sum_\alpha q_\alpha = k$. The left-right order of $i_\beta$ and $i_{\beta+1}$ in Eq.\ \eqref{eq:hord} depends on whether the gap corresponds to a $\bm J$ or a $\bm J^\top$ (Fig.\ \ref{fig:defn}). Under this ordering, neighboring gaps share an index: $(H^\alpha_n)_2 = (H^\alpha_{n+1})_1$, and the boundary terms $H^\alpha_1, H^\alpha_{q_\alpha}$ contain a fixed $a_\mu$ index as their first and second entry, respectively (Sec.\ \ref{sec:ordlemma}). The fact that the $J_{ij}$ gaps may be ordered such that neighboring $J_{ij}$ gaps share an index organizes the expansion into repeated matrix products of $\bm J$ and $\bm J^\top$ as illustrated below.

We now translate a few diagrams into their corresponding equations, starting with a diagram in which there are no bubbles and no connections beyond the horizontal pairings between neighbors. We call these ``chain diagrams," and they  correspond to index vectors $\mathbf{i,j}$ in which each index appears exactly twice. For example, the following chain diagram:
\begin{center}
    \centering
    \includegraphics[width=\linewidth]{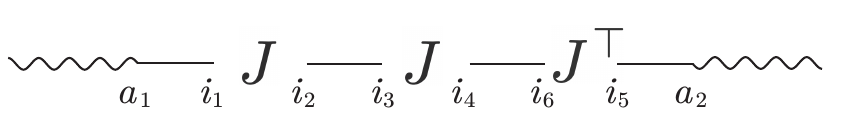}
\end{center}
\noindent corresponds to the index constraints $(a_1 = i_1),  (i_2 =i_3), (i_4 = i_6), (i_5 = a_2)$ with all other indices distinct (the parentheses are for readability). The diagram therefore represents the sum
\begin{gather}
\begin{aligned}
i^3\sum_{p_1 \neq p_2  \neq (a_1, a_2)} 
&\langle f_1 \hat x_{a_1} \rangle_0 J_{a_1 p_1} 
\langle \phi_{p_1} \hat x_{p_1}\rangle_0 J_{p_1 p_2} 
\\
&\langle \phi_{p_2} \phi_{p_2} \rangle_0 
J_{ a_2 p_2} 
\langle  \hat x_{a_2} f_2\rangle_0, 
\end{aligned}
\label{eq:chainex}
\end{gather}
where the factor $1/3!$ was canceled by the $\symfactor=3!$ inequivalent ways to permute the $J_{ij}$ gaps, and the sum is over $p_1, p_2$. 

The above sum has the form of a matrix product $[\bm J \bm J \bm J^\top]_{a_1 a_2}$, with the restriction that the contracted indices, $p_1, p_2$, are distinct from each other and from the $a_1,a_2$. That is, the indices do not repeat more than twice. As a result of this constraint, the only correlation functions appearing in Eq.\ \eqref{eq:chainex} are two-point functions. 

To represent terms in which indices repeat additional times, one draws arcs.  For example, the diagram 
\begin{center}
    \includegraphics[width=\linewidth]{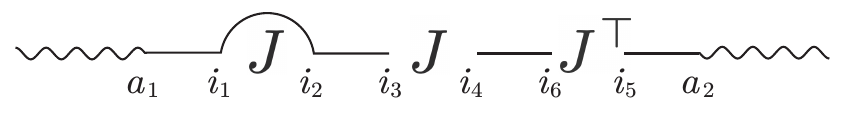}
\end{center}
\noindent yields the index constraints $(a_1 = i_1 = i_2 =i_3)$, $(i_4=i_6)$, and $(i_5=a_2)$, with all other indices distinct. The corresponding sum is
\begin{gather}
\begin{aligned}
    i^3 \sum_{p \neq (a_1, a_2)}
    &\langle f_1 \hat x_{a_1} \phi_{a_1} \hat x_{a_1}\rangle_0 
    J_{a_1 a_1} J_{a_1 p} J_{a_2 p} 
    \\
    &\langle \phi_p \phi_p\rangle_0  
    \langle \hat x_{a_2} f_2 \rangle_0.
\end{aligned}
\end{gather}
Note that the presence of four repeated indices leads to a four point correlation function in the $\langle \cdot \rangle_0$ average. Moreover, the extra identification, $i_1=i_2$, implies that there is one fewer free index to sum over. The diagram is therefore suppressed by a factor of $N^{-1/2}$ relative to the preceding chain diagram as shown in Sec.\ \ref{sec:vardiags}.

From the last example, we might expect that any additional index constraints beyond the pairing of neighboring $J_{ij}$ gaps leads to a large-$N$ suppression of the diagram. This is not the case. For example, consider the ``rainbow" 
\begin{center}
    \includegraphics[width=\linewidth]{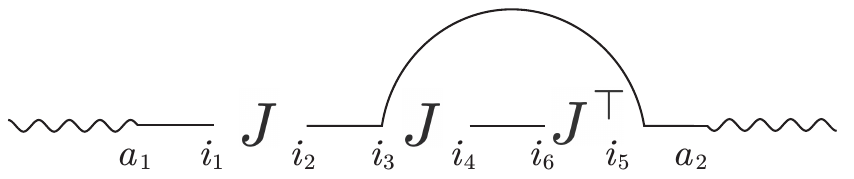}
\end{center}

\noindent This diagram has the index constraints $(a_1 = i_1),  (i_2 = i_3=i_5 = a_2),$ and $(i_4 = i_6)$, with the rest distinct. This gives the sum
\begin{gather}
\begin{aligned}
    \frac {i^3} 2 \sum_{p \neq (a_1, a_2)}
    &\langle f_1 \hat x_{a_1}\rangle_0 J_{a_1 a_2} 
    \langle \phi_{a_2} \hat x_{a_2} \hat x_{a_2} f_2\rangle_0  
    \\
    &J_{a_2p}^2 \langle \phi_p \phi_p\rangle_0   .
\end{aligned}
    \label{eq:rainbowex}
\end{gather}
Note that $\symfactor/k!l!$ in this case leaves behind a factor of $1/2$, as one of the permutations---the swapping of the two $J_{ij}$ gaps connected by the arcs, $(i_3, i_4) \leftrightarrow (i_5, i_6)$---does not change the pattern of index connections. Again we see that the additional index constraints create a four-point function over $\phi$ and  $\hat x$. 

In the above diagram, the summation over indices once again has the form of a constrained matrix product $[\bm J \bm J \bm J^\top]_{a_1 a_2}$, but we now restrict $\bm J \bm J^\top$ to its diagonal entries. The diagonal entries of $\bm J \bm J^\top$ behave as $[\bm J \bm J^\top]_{aa} = \sigma^2 + O(N^{-1/2})$. As such, the extra index connection between $\bm J$ and $\bm J^\top$ does not suppress the large-$N$ scaling of the sum. We will show below that such $\bm J, \bm J^\top$ pairings are in fact the only types of index connections beyond those between neighboring $J_{ij}$ gaps that do not suppress the large-$N$ behavior of a diagram. 

As a final example, consider the following diagram containing a $\Delta$ bubble: 
\begin{center}
    \includegraphics[width=\linewidth]{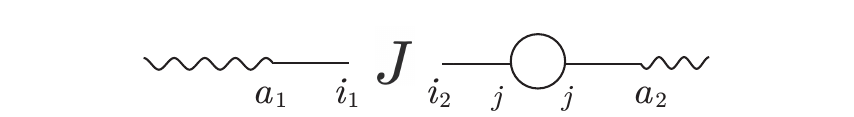}   
\end{center}
\noindent This yields the pattern of index constraints $(a_1=i_1)$ and $(i_2 = j = a_2)$. Translating back to equations: 
\begin{gather}
    \frac i  2 \langle f_1 \hat x_{a_1}\rangle_0 J_{a_1 a_2} \langle \phi_{a_2} \hat x_{a_2} \hat x_{a_2} f_{2} \rangle_0 \Delta,
\end{gather}
where the $1/2$ factor comes from the definition of the bubble (Fig.\ \ref{fig:defn}, Eq.\ \eqref{eq:series}). Note the similarity to the preceding example, Eq.\ \eqref{eq:rainbowex}. Indeed, making the approximation $\sum_{p \neq (a_1, a_2)} J_{a_2 p}^2 = \sigma^2 + O(N^{-1/2})$ and using the mean-field condition, $\Delta = \sigma^2 \langle \phi_j \phi_j \rangle_0$ (Eq.\ \eqref{eq:dmft-selfcons}), we can see that the two diagrams are sign-flipped versions of each other. That is, the rainbow diagram above and the diagram obtained by replacing the $\bm J \bm J^\top$ pairing with a $\Delta$ bubble cancel each other. We will see in Sec.\ \ref{sec:cancellation} that this mechanism, together with the fact that bubbles may be inserted along any edge, cancels all diagrams besides the chains to leading order in $1/\sqrt{N}$. 

At a given order $k,l$ in Eq.\ \eqref{eq:series}, one sums over every distinct way to connect the indices while obeying the index constraints on the fixed indices, $a_\mu$. For example at the order $k=3, l=0$ with two fixed indices $a_1\neq a_2$, the nonzero terms in the series Eq.\ \eqref{eq:series} generate the diagrams
\begin{center}
    \includegraphics[width=\linewidth]{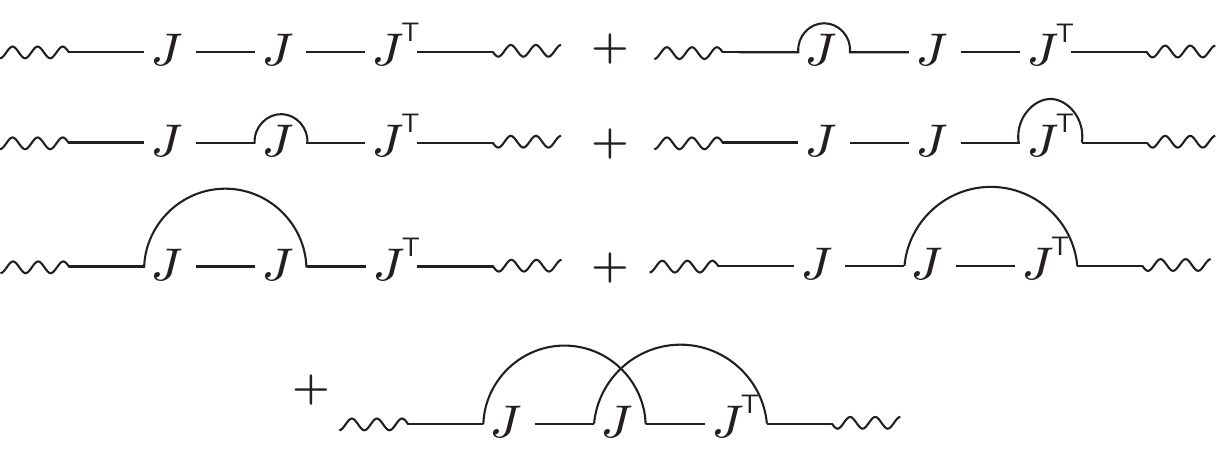}
\end{center}
\noindent Note that the arc connecting the outermost legs of the first and the last $J_{ij}$ gaps cannot be drawn, as this would imply $a_1=a_2.$ The remaining diagrams at this order $k=3, l=0,$ can be obtained by exchanging $\bm J$ variables with $\bm J^\top$ and vice-versa. 

\begin{figure*}
    \centering
    \includegraphics[width=0.99\textwidth]{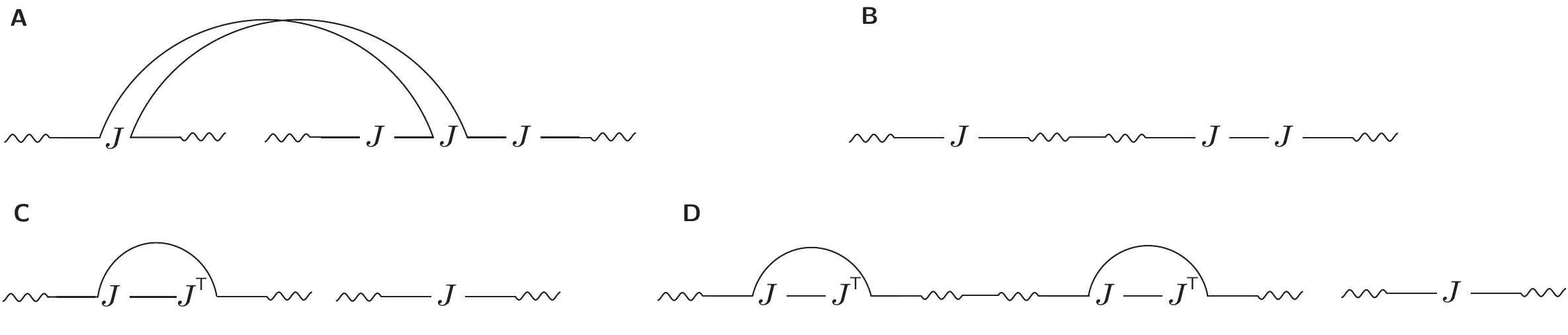}
    \caption{(A-B) Non-maximally disconnected diagrams. (C-D) Maximally disconnected, leading order diagrams. Note that both (B) and (C) contain four wavy lines and a set of repeated fixed indices, $a_\mu=a_\nu$. The diagram in C contains 2 disconnected blocks, while the diagram in B contains only one.}
    \label{fig:max-disc}
\end{figure*}

\section{Sketch of the proof}
\label{sec:proof-sketch}
We now sketch the proof of Eqs.\ \eqref{eq:introshenhu}, \eqref{eq:responseintro}, and \eqref{eq:simplemoment}. The strategy is to identify the diagrams that survive at leading nontrivial order in $1/\sqrt N$, show that diagrams involving higher order correlation functions systematically cancel each other, and resum the remaining series. The result is a closed-form solution for moments and response functions of intensive order.

The first step is to identify the leading order diagrams. Since $\bm J$ is held fixed, this cannot be done by averaging the diagrams over $\bm J$. Instead, for a typical realization of $\bm J$, we estimate the magnitude of a diagram $D$ through its mean square $\langle |D|^2\rangle_{\bm J}$ (Sec.\ \ref{sec:vardiags}). This turns the large-$N$ problem into a counting problem: each summed index brings a factor of $N$, while each $J_{ij}$ gap brings a factor of $N^{-1/2}$ to the average $\langle |D|^2\rangle_{\bm J}$. The leading diagrams are therefore those maximizing the number of summed indices at a given order of the expansion. This argument shows that the leading order diagrams have the form of the chains and rainbows shown in the examples above with no additional connections drawn between the $m/2$ blocks. The latter condition corresponds to an effective Wick factorization of the moments. The subleading diagrams are suppressed by a factor of $N^{-1/2}$ relative to the diagrams that are retained. 

What types of sums do the remaining diagrams correspond to? The chain diagrams are constrained products of $\bm J$ and $\bm J^\top$ in which each contracted index appears no more than twice. As a result, every average $\langle \cdot\rangle_0$ appearing in a chain is a two-point function. On the other hand, the rainbow diagrams involve products of $\bm J$ and $\bm J^\top$ with the diagonal entries of $\bm{J J}^\top$. As such, they lead to a greater number of repeated indices and generate four- and higher-point functions  in the $\langle \cdot \rangle_0$ average. Diagrams involving $\Delta$ bubbles similarly generate higher correlation functions. 

In Sec.\ \ref{sec:cancellation}, we show that the rainbow diagrams and those containing $\Delta$ bubbles systematically cancel one another to leading nontrivial order in $N^{-1/2}$. This leaves only the chains. The expansion of a correlation function $\langle f_1 \cdots f_m\rangle$ for all indices distinct can therefore be expressed purely in terms of two-point functions.

The final step is to resum the remaining series. The difficulty at this stage is that the sum generated by the chain diagrams involves matrix products in which indices are constrained to repeat as little as possible. The key step is the following lemma, stated precisely in Sec.\ \ref{sec:resum}. Given a random matrix $\bm J$ with i.i.d.\ elements and two positive integers, $p,q\geq 1$, we show that 
\begin{gather}
\begin{aligned}
    \bm J^p (\bm J^\top)^q 
    &= (\bm J^p (\bm J^\top)^q)^{\mathrm{no-repeats}} 
    + \sigma ^2 \bm J^{p-1} (\bm J^\top)^{q-1} 
    \\
    &\quad + O(N^{-1/2} \bm I + N^{-1} \bm 1),
\end{aligned}
    \label{eq:sketchjlemma}
\end{gather}
where we define $(\cdot)^{\mathrm{no-repeats}}$ to be the matrix product in which indices repeat as little as possible. For example, for two matrices $A,B$, we have $(AB)^{\mathrm{no-repeats}}_{ij} = \sum_{k \neq (i,j)} A_{ik} B_{kj}$, and we define $[AB]^{\mathrm{no-repeats}}_{ii} = 0$ to avoid double counting certain terms (see Eq.\ \eqref{eq:no-repeats-defn} for the full definition). The above lemma states that the leading order behavior of a random matrix product follows from terms in which indices repeat as little as possible, with the exception of the diagonal terms in a $\bm J \bm J^\top$ product, which bring a factor of $\sigma ^2 \bm I +O(N^{-1/2}\bm I)$. This allows us to resum the no-repeats series generated by the chain diagrams, which in turn establishes the Shen-Hu conjecture and equations for the linear response functions, $\bm R^\phi, \bm R^x$. Summing over higher correlation functions then establishes the generalized Wick formula to leading non-trivial order in $1/\sqrt N$.

\section{Proof}
\label{sec:proof}

\subsection{Leading order diagrams}
\label{sec:leadingorddefn}

In this section we characterize the leading order diagrams in the $1/\sqrt N$ expansion as maximally disconnected diagrams in which the only arcs form pairings between $\bm J$ and $\bm J^\top$, which we call rainbows. 

For a given choice of fixed indices $a_1, \cdots, a_m$, the leading order diagrams are maximally disconnected across blocks. We say that two blocks $H^\alpha, H^\beta$ are connected if they have a shared index, and we define $Q$ as the number of disconnected blocks in a diagram. A diagram $ D $ is \emph{maximally disconnected} if $Q$ attains its maximum among all the diagrams in the expansion of $\langle f_1 \cdots f_m \rangle$ so that $Q=Q_{\mathrm{max}}  $ (Fig.\ \ref{fig:max-disc}). 

When the $a_\mu$ are all distinct $a_\mu \neq a_\nu$, maximally disconnected diagrams are simply those in which no additional index connections (i.e. no arcs) exist between the $ m/2 $ blocks, and $Q_{\max} = m/2 $.
When the $a_\mu$ repeat, any maximally disconnected diagram can be joined along the shared $a_\mu$ indices into a set of disconnected \emph{coarse blocks,} $\bar H^\alpha,$ that satisfy the ``neighbors share an index" condition $(\bar H^\alpha_n)_2 = (\bar H^\alpha_{n+1})_1$ (Fig.\ \ref{fig:max-disc}D; see Appendix Sec.\ \ref{sec:coarse} for a proof). We will see that the large-$N$ dominance of maximally disconnected diagrams corresponds to a Wick factorization of the moments.

\begin{figure}
    \centering
    \includegraphics[width=0.95\linewidth]{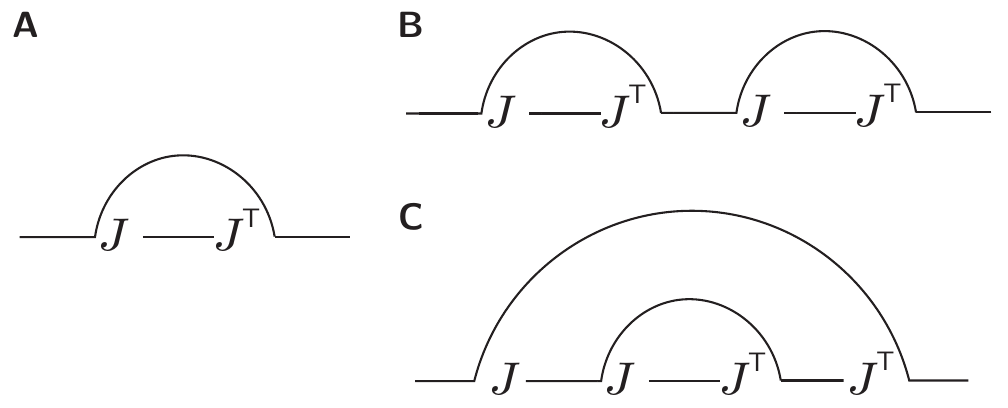}
   \caption{(A) Bare rainbow. (B) Join and (C) nest operations applied to the bare rainbow.}
    \label{fig:rainbow}
\end{figure}

In leading order diagrams, the connections between $J_{ij}$ gaps in a diagram must form \emph{rainbows.} Starting from the ``bare rainbow" in Fig.\ \ref{fig:rainbow}A, we recursively define a rainbow as any sequence of $J_{ij}$ gaps formed by taking a pre-existing rainbow and either connecting it with a neighboring rainbow or nesting the rainbow under an arc connecting a $\bm J$ to a $\bm J^\top$ so that a factor of $J_{ij}^2$ forms. We define rainbows to be sequences of $J_{ij}$ gaps formed by successive applications of these two operations, without additional arcs.

\begin{figure*}[t]
    \centering
    \includegraphics[width=0.75\textwidth]{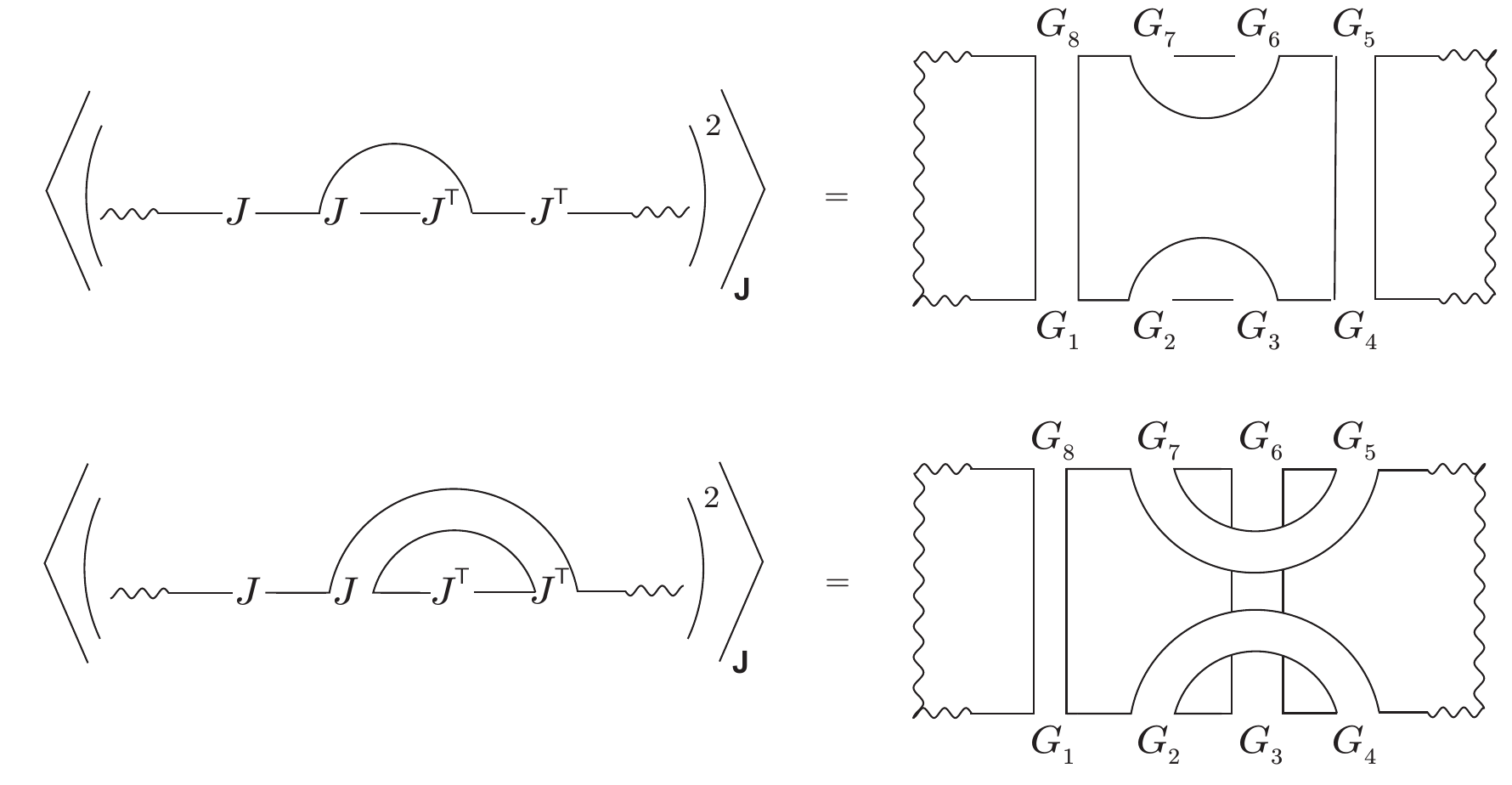}
    \caption{Mean-square diagrams. We suppress the $J$ symbols so we can label the gaps $G_1, \cdots, G_8$.}
    \label{fig:vars}
\end{figure*}

Note that the bare rainbow connecting a $\bm J^\top$ on the left to a $\bm J$ on the right produces a factor of $\langle \hat x \hat x\rangle_0  = 0$ (Eq.\ \eqref{eq:xhatvanish}). In fact, any rainbow in which a $\bm J^\top$ is paired to a $\bm J$ to its right is zero. This can be shown by induction on the number of joins and nests (Fig.\ \ref{fig:rainbow}). Assuming any rainbow containing a $\bm J^\top \bm J$ pairing formed from $K$ nests and joins is zero, consider a rainbow formed from $K+1$ such operations. If the $K+1$ step involves a join, then there are no new pairings of $\bm J, \bm J^\top$. Thus, at least one of the joined rainbows contains a $\bm J^\top \bm J$ pairing. Whichever $\langle \cdot\rangle_0$ average caused that rainbow to be zero remains. If on the other hand the $K+1$ step is a nest, then the outgoing legs of the rainbow being nested must carry $\hat x$ fields, otherwise it is zero by the inductive hypothesis. Thus, if the $K+1$ step is a nest in which a $\bm J^\top$ on the left pairs to a $\bm J$ on the right, then a factor of $\langle (\hat x_i \hat x_i)^b\rangle_0 = 0$ must form for some integer $b$ (Eq.\ \eqref{eq:xhatvanish}). The only non-zero rainbow diagrams are therefore those consisting of pairings of the form $\bm J \bm J^\top$.

We show in the next section that for leading order diagrams, the only index identifications beyond the connections between neighboring $J_{ij}$ gaps $H^\alpha_n$ and $H^\alpha_{n+1}$ (i.e. the only arcs) must come from rainbow pairings. Any diagram featuring index connections not described by these two sources is subleading (e.g.\ the second example in Sec.\ \ref{sec:defns}). 

The leading order diagrams consist of sequences of rainbows and chains, together with the diagrams obtained by arbitrary insertions of $\Delta$ bubbles into the edges. Importantly, for blocks in which the two fixed indices $a_\mu$ are distinct, the leading order diagrams may consist of mixtures of chains and rainbows, while for blocks in which both fixed indices are equal, $a_\mu=a_\nu$ there can be no chains between the wavy lines (Fig.\ \ref{fig:max-disc}C-D). That is, the only structure that can exist between two connected wavy lines is a rainbow. The reason is that the identification $a_\mu=a_\nu$ is not implied by the neighbor rule on $\bar H^\alpha$. As such it must come from a rainbow pairing. These statements are proved in the next section.

\subsection{Large-N analysis}
\label{sec:vardiags}

We show in this section that leading order diagrams satisfy the two properties defined above: 
\begin{enumerate}
    \item The blocks are maximally disconnected. 
    \item The only index identifications beyond those enforced by the neighbor condition on $\bar H^\alpha$ are rainbow pairings.
\end{enumerate}

\noindent For a typical draw of $\bm J$, the magnitude of a diagram $D$ is of order $\sqrt{\langle |D|^2\rangle_{\bm J}}$. We keep only those $D$ whose mean square is of leading order. To do this, we represent with diagrams the various terms appearing in the mean square: 
\begin{align}
    \langle |D|^2 \rangle_{\J} 
    &= 
    \bigg(\frac{ \symfactor }{2^l k! l!} \bigg)^2
    \sum_{\substack{(\mathbf i, \mathbf j ) \in \mathcal A 
    \\
    (\mathbf i', \mathbf j')\in \mathcal A
    } }
    \nonumber \\
    &\quad
    \big\langle 
    \langle f_1 \cdots f_m 
    \hat x_{i_1} J_{i_1 i_2} \phi_{i_2}
    \cdots \hat x_{j_l} \Delta \hat x_{j_l}
    \rangle_0
    \nonumber \\
    &\qquad
     \langle f_1 \cdots f_m   
     \hat x_{i'_1} J_{i'_1 i'_2} \phi_{i'_2}
    \cdots \hat x_{j'_l} \Delta \hat x_{j'_l}
    \rangle_0\big\rangle_{\J}
    \label{eq:vdiags}
\end{align}
Graphically, we draw two copies of the diagram $D$ and sum over the various ways to connect the indices $(\mathbf i, \mathbf j )$ and $(\mathbf i', \mathbf j')$ that obey the index constraints in the original diagram (Fig.\ \ref{fig:vars}).

The blocks of a mean-square diagram consist of two copies of the  $H^\alpha$ blocks in $D$. We label the corresponding indices of the $J_{ij}$ gaps in a cyclic order as
\begin{gather}
    G^\alpha = (H^\alpha_1, \cdots, H^\alpha_{q_\alpha}, H^{\alpha '}_{q_\alpha}, \cdots, H^{\alpha '}_1), 
    \label{eq:g-const}
\end{gather}
where $\alpha = 1, \cdots, m/2$, and the second copy $H^{\alpha '}_n$ has its indices reversed (Fig.\ \ref{fig:vars}). Using the fact that the indices $a_\mu$ are fixed, this cyclic ordering implies that boundary components must share an index: $(G^\alpha_1)_1 = (G^\alpha_{2q_\alpha})_2$ and $(G^\alpha_{q_\alpha})_2 = (G^\alpha_{q_\alpha+1})_1$  (Fig.\ \ref{fig:vars}). Note too that the number of disconnected $G^\alpha$ blocks $Q'$ verifies $Q'\leq Q$ with $Q$ the number of disconnected $H^\alpha$.

To start the large-$N$ analysis, note that by Eq.\ \eqref{eq:moment}, a mean-square diagram $V$ scales as 
\begin{align}
    V
    &=
    \sum_{(\mathbf i, \mathbf j), (\mathbf i', \mathbf j') \in \mathcal B}
    \big\langle 
    \langle f_1 \cdots f_m 
    \hat x_{i_1} J_{i_1 i_2} \phi_{i_2}
    \cdots \hat x_{j_l} \Delta \hat x_{j_l}
    \rangle_0
    \nonumber \\
    &\qquad
     \langle f_1 \cdots f_m   
     \hat x_{i'_1} J_{i'_1 i'_2} \phi_{i'_2}
    \cdots \hat x_{j'_l} \Delta \hat x_{j'_l}
    \rangle_0\big\rangle_{\J}
    \nonumber \\
    & \leq   
    \sum_{(\mathbf i, \mathbf j), (\mathbf i', \mathbf j') \in \mathcal B} 
    C' N^{-k},
\label{eq:vdiagscale}
\end{align}
where the constant factor $C'$ coming from the time integrals and permutation symmetries $\symfactor\leq k! l! $ does not scale with $N$, and $\mathcal B$ is the set of indices obeying the constraints of the diagram. Since the $j$ index attached to a $\Delta$ bubble must be connected to an index from a $J_{ij}$ gap or wavy line, the insertion of a $\Delta$ bubble can never introduce an additional free index to sum over. We therefore focus on the case without bubbles ($l=0$) without a loss of generality.

In what follows, we first adapt a graphical argument from random matrix theory to show that non-maximally disconnected diagrams and mean-square diagrams involving higher moments, $\langle J_{ij}^q\rangle_{\bm J}$ for $q>2$, are both subleading (see for example Ref.\ \cite{anderson2010introduction} Ch. 2). This leaves maximally disconnected diagrams involving second moments of $J_{ij}$. Embedding the remaining mean-square diagrams as ribbon graphs and applying Euler's theorem (see e.g.\ Ref.\ \cite{bessis1980quantum}) then shows that the chains and rainbows are the only diagrams $D$ capable of generating leading order mean-square diagrams $\langle |D|^2\rangle_{\bm J}$.

\begin{figure}[t]
    \centering
    \includegraphics[width=0.7\linewidth]{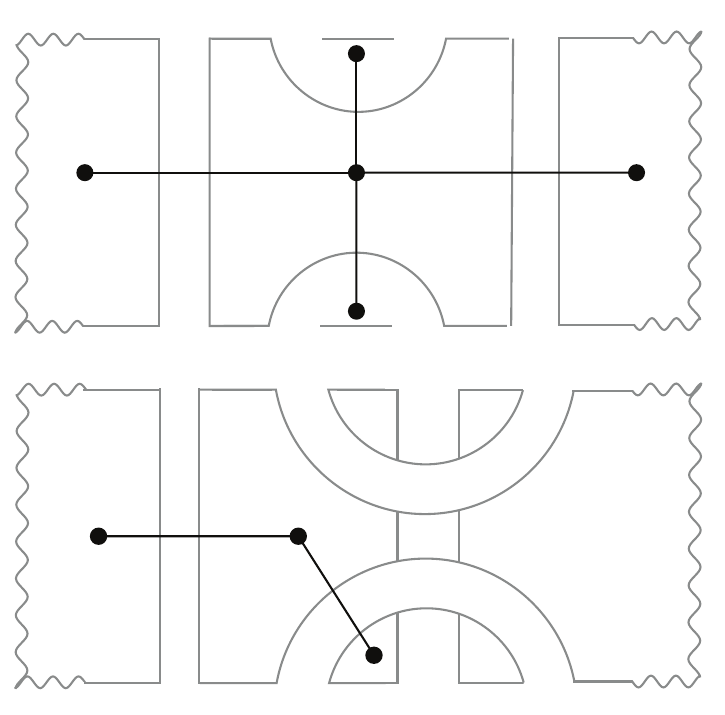}
    \caption{Multigraphs for the mean-square diagrams in Fig.\ \ref{fig:vars}. We can see that the diagram on the right has fewer vertices and is therefore subleading.}
    \label{fig:multigraph}
\end{figure}

To start, let us define a multigraph $\mathcal M$ consisting of $L$ vertices, one for each unique index in the mean-square diagram, $\langle |D|^2\rangle_{\bm J}$. That is, each unique left and right index of the $J_{ij}$ gaps corresponds to a vertex in $\mathcal M$ (Fig.\ \ref{fig:multigraph}). It follows that Eq.\ \eqref{eq:vdiagscale} obeys
\begin{gather}
    \sum_{(\mathbf i, \mathbf j), (\mathbf i', \mathbf j') \in \mathcal B} C N^{-k}= O( N^{L - k -r}), 
    \label{eq:graphscaling}
\end{gather}
where the $-r$ factor comes from the fact that the $r$ distinct indices $a_\mu$ are held fixed. 

Next, define a partition $P$ of the $J_{ij}$ gaps by grouping together powers of $J_{ij}$. That is, each $p\in P$ consists of index pairs, $G^\alpha_n$, that have equal left and right $J_{ij}$ indices. The condition $\langle J_{ij} \rangle_{\bm J} = 0$, implies that for every non-zero-mean-square diagram, $|p|\geq 2$ for all $p \in P$, and $|P| \leq k$ with equality if and only if all moments are second order, where we use $|p|$ to denote the number of elements of the set $p$.
Using this partition $P$, we draw an edge between the vertices in $\mathcal M$ corresponding to the left and right indices of each $p\in P$ (Fig.\ \ref{fig:multigraph}). It follows that $|P| = E$, where $E$ is the number of edges in the multigraph $\mathcal M$. 

We now prove that 
\begin{gather}
    L \leq |P| + Q' \leq k + Q_{\max} .
    \label{eq:connectedgraph}
\end{gather}
We first show that the subgraph corresponding to each of the $m/2$ blocks in $G^\alpha$ is connected (Fig.\ \ref{fig:multigraph}). To see this, consider an arbitrary block $G^\alpha$. For each $G^\alpha_n$, traverse the edge between the left and right vertices corresponding to the first and second indices in $G^\alpha_n$. The fact that adjacent entries in $G^\alpha$ share an index, together with the boundary conditions on the block, imply that the resulting walk traverses the whole subgraph. Each of the $m/2$ subgraphs are therefore connected. Moreover, any two subgraphs sharing an index must have a common vertex in $\mathcal M$ and are thus connected. Denoting the number of vertices in the $i$th such disconnected subgraph by $L_i$, the connectedness of the subgraph implies $L_i \leq E_i + 1$. Summing over $i$ establishes Eq.\ \eqref{eq:connectedgraph}. 

It follows that mean-square diagrams with $Q'<Q_{\mathrm{max}}$ and those involving higher order moments, $|P|<k$, are subleading. It remains to show that among the maximally disconnected $D$ featuring pairwise $J_{ij}$ connections, only the rainbows and chains defined in the previous section survive. 

Consider the mean-square diagrams generated by a maximally disconnected diagram $D$. Recall that the $m/2$ blocks, $H^\alpha_n,$ can be grouped into mutually disconnected coarse blocks, $\bar H^\alpha_n$ (Appendix Sec.\ \ref{sec:coarse}). Thus, in a maximally disconnected variance diagram, $\langle |D|^2\rangle_{\bm J}$, we may group the $G^\alpha_n$ into a set of mutually disconnected coarse blocks, $\bar G^\alpha_n$ by following Eq.\ \eqref{eq:g-const}, replacing $H^\alpha_n$ with $\bar H^\alpha_n$. We now establish the claim by representing these coarse blocks as ribbon graphs and applying well known arguments involving planar diagrams \cite{bessis1980quantum}.

\begin{figure}[t]
    \includegraphics[width=\linewidth]{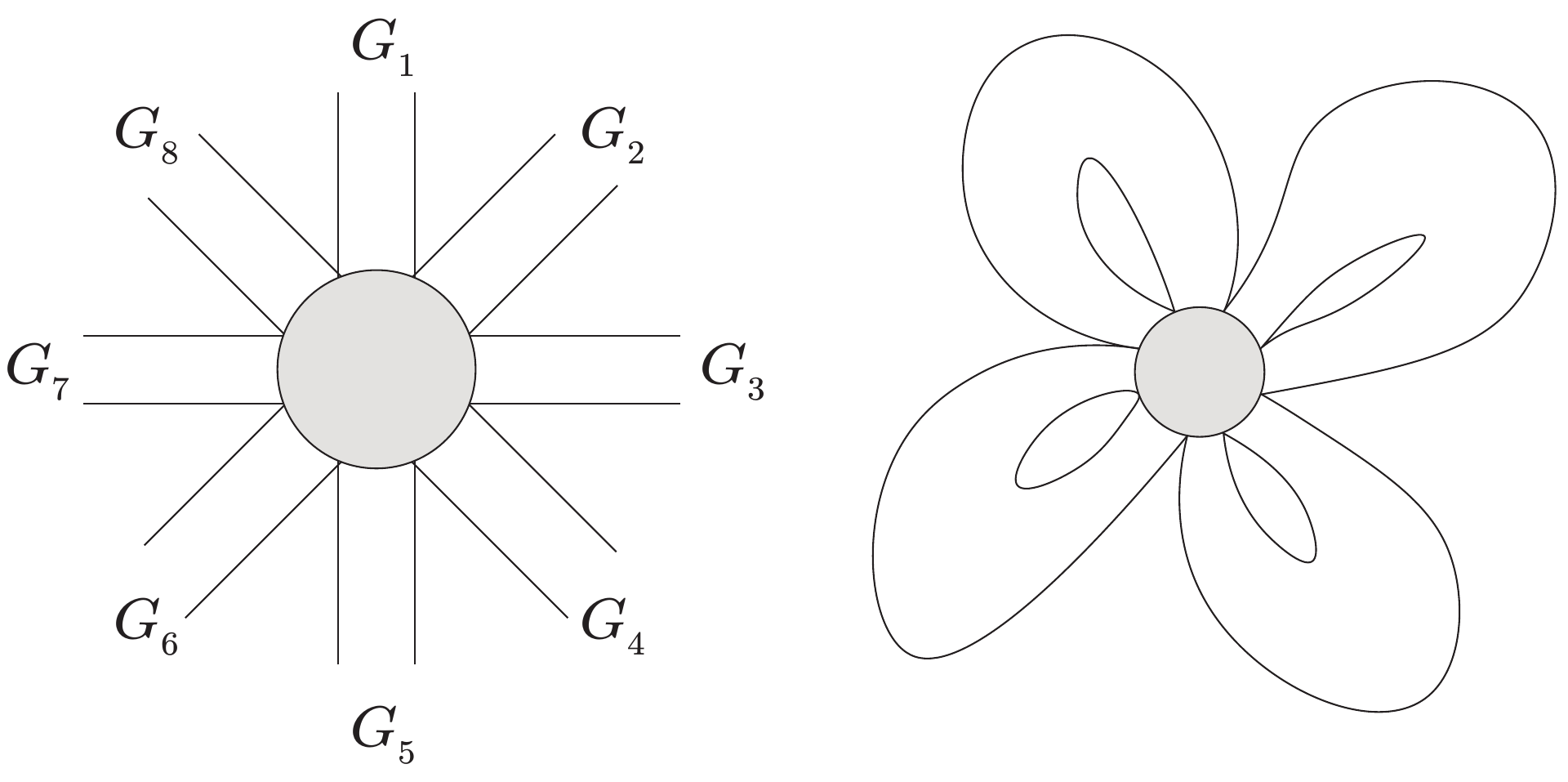}
    \caption{Ribbon graph for the top diagram in Fig.\ \ref{fig:vars}.}
    \label{fig:ribbon}
\end{figure}

For each coarse block, we draw a single vertex and attach a half edge for each $\bar G^\alpha_n$, following the cyclic order in $\bar G^\alpha$ (Fig.\ \ref{fig:ribbon}; Ref.\ \cite{bessis1980quantum}). Half edges are then joined according to the $J_{ij}$ pairings. If no more index identifications exist beyond the neighbor conditions in $\bar G^\alpha$ and the pairings represented by the partition $P$, then the number of distinct indices in each block, $L_\alpha$, is the number of faces in the ribbon graph, $L_\alpha = F_\alpha$. If additional index identifications exist, it follows that $L_\alpha < F_\alpha$. Euler's theorem then implies
\begin{gather}
    F_\alpha = 1 + E_\alpha - g, 
    \label{eq:euler}
\end{gather}
where $g$ is the (non-orientable) genus of the surface, and $E_\alpha$ is the number of edges in the ribbon graph. The use of the non-orientable genus accounts for the possibility of ``twisted" edges in which two $\bm J$ (or two $\bm J^\top$) are paired with each other. The number of edges $E_\alpha$ is simply half the number of $J_{ij}$ gaps in $\bar G^\alpha$ so that $\sum_\alpha E_\alpha = k$. Summing over ribbons  therefore gives $L \leq k + Q_{max}$ with equality if and only if the diagram is planar and has no additional index constraints beyond those of the cyclic ordering and the pairings in $P$. 

Finally, we show that the planar pairings correspond to the rainbows and chains. Consider first ribbon graphs in which there exists a pairing between two $J_{ij}$ gaps, $\bar G^\alpha_{n}, \bar G^\alpha_m$ with $1\leq n < m\leq q_\alpha$. To avoid a crossing of lines, the half edges corresponding to all $\bar G^\alpha_s$ with $n < s <m$ must be paired amongst themselves. Proceeding by induction, there must exist at least one $b$ with $n\leq b< m$ such that $\bar G^\alpha_b, \bar G^\alpha_{b+1}$ are paired. Any additional pairings among the $\bar G^\alpha_s$ must arise from nesting such a pair under arcs or pairing other neighboring $\bar G^\alpha_s$. Since twisted edges increase the genus $g$ in Eq.\ \eqref{eq:euler}, these must be pairings between $\bm J$ and $\bm J^\top$ gaps. These are rainbows. Conversely, pairings between the first $q_\alpha$ and the last $q_\alpha$ half edges in $\bar G^\alpha$ can only be generated by chains in $D$. This establishes the claim.

\subsection{Cancellation of rainbows and bubbles}
\label{sec:cancellation}
We now show that the rainbow diagrams and those obtained by replacing the ``bare rainbow" (Fig.\ \ref{fig:rainbow}) with a $\Delta$ bubble cancel one another out. This leaves only the diagrams without any paired $J_{ij}$ gaps---i.e. the chain diagrams and the bare terms $k=l=0$. 

Ignoring combinatorial coefficients for the moment, the bare rainbow contributes a factor 
\begin{gather}
    i^2 \sum_{j \in \mathcal C} J^2_{ij} \hat x_i \langle \phi_j\phi_j \rangle_0 \hat x_i = - \hat x_i \Delta \hat x_i   + O(N^{-1/2})
\end{gather}
 where the sum over $j$ is restricted to some $O(N)$ subset of indices $\mathcal C$ that depends on the rest of the diagram, and we have used Eq.\ \eqref{eq:dmft-selfcons}. The $O(N^{-1/2})$ residual is given by a term proportional to $(\sum_{j\in \mathcal C}J_{ij}^2 - \sigma^2)$. Since the $j$ indices are constrained to be distinct from all other indices, this residual is explicitly uncorrelated with the other terms in the diagram. Neglecting it therefore brings an error that is suppressed by a factor of $N^{-1/2}$ relative to the rest of the diagram. Up to the coefficients $\symfactor/(k! l! 2^l)$ and to leading non-trivial order in $N$, the bare rainbows and the bubbles are therefore sign flipped versions of one another. 

It remains to show that the combinatorics verify the cancellation. We give a heuristic argument here and defer the rigorous proof to the Appendix Sec.\ \ref{sec:cancel-comb}. We first reparametrize the problem slightly by writing the combinatorial coefficient as $\symfactor = k!l!/\symsymbol$, where $\symsymbol$ denotes the number of permutations that leaves the pattern of equal indices in a diagram unchanged. For example, in the chain diagram of Sec.\ \ref{sec:defns}, only the identity mapping leaves the index equalities unchanged, $\symsymbol=1$, while in the rainbow diagram, the permutation $(i_3, i_4) \leftrightarrow (i_5, i_6)$ does as well, leading to $\symsymbol=2$. More generally, for a rainbow with $p$ pairs there are $2^p$ such symmetries. 

The additional factor of $1/2$  associated with each pair cancels the $1/2^l$ factor associated with $l$-bubbles. That is, starting from a rainbow diagram, the sum over all diagrams obtained by replacing bare rainbows by bubbles is zero. For example, the following diagrams in the expansion of the two-point function $\langle x_ax_a\rangle$ cancel, so that up to $O(1)$ 
\begin{center}
    \includegraphics[width=\linewidth]{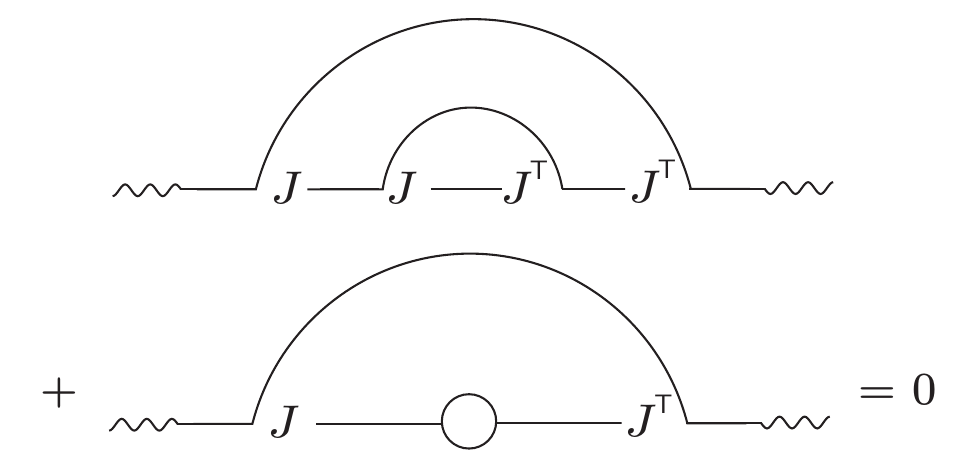}
\end{center}
\noindent More generally the symmetries between rainbow pairs cancel the $1/2^l$ factors introduced by the bubbles. We give a rigorous proof of this fact taking into account additional symmetries of the diagrams in the Appendix Sec.\ \ref{sec:cancel-comb}. 

What is left after this cancellation? Consider first a coarse block $\bar H^\alpha$ in which all wavy lines correspond to the same fixed index (e.g.\ the diagram above and the first block of Fig.\ \ref{fig:max-disc}D). We have shown in the preceding sections that to leading order in $1/\sqrt{N}$, the only type of graph that can exist between such wavy lines are rainbows. All such diagrams cancel with the diagrams obtained by replacing bare rainbows with $\Delta$ bubbles. Thus, the only type of block $\bar H^\alpha$ with a single unique fixed index that survives is the bare term $k=l=0$.  

\begin{figure}[t]
    \centering
    \includegraphics[width=0.90\linewidth]{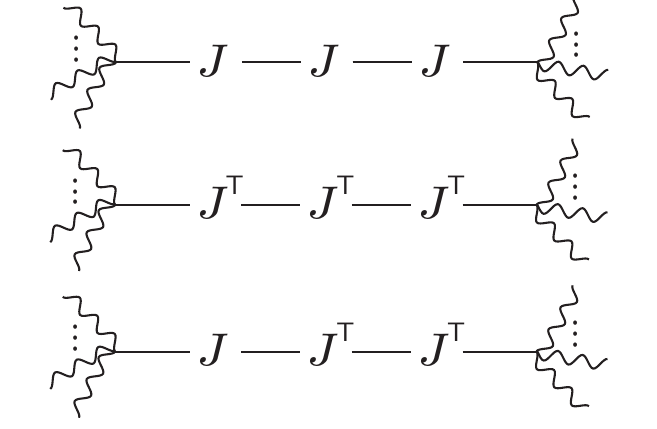}
    \caption{Example leading order chain diagrams.}
    \label{fig:chains}
\end{figure}

Consider then ``mixed blocks" $\bar H^\alpha_n$ in which there are two unique fixed indices, each repeating an odd number of times within the block (e.g.\ the second block of Fig.\ \ref{fig:max-disc}D). Between any two wavy lines corresponding to the same index, the only possible structures are again rainbows. Cancelling the rainbows leaves only a ``tuft" of bare terms on either side of the block corresponding to a factor of $ \prod_{\mu : a_\mu = \bar a} f_\mu(x_{a_\mu}, \hat x_{a_\mu}),$ where $\bar a$ is the shared fixed index (Fig.\ \ref{fig:chains}). 

On the other hand, between two wavy lines corresponding to \emph{distinct} indices, the bare term $\langle f_\mu f_\nu \rangle_0$ is zero, and the remaining diagrams are chains in which each summed index appears exactly twice. Given that $\langle \hat x \hat x\rangle_0 =0$, it follows that no factor of $\bm J^\top \bm J$ may form in a chain diagram. Thus, the chains have the form of repeated powers, $\bm J^{p} (\bm J^\top)^{q}$ (Fig.\ \ref{fig:chains}).

\subsection{Resummation and no-repeats lemma}
\label{sec:resum}
In this section we resum the remaining diagrams in the expansion of $\langle \ob_1 \cdots \ob_m\rangle$. 

We begin by considering the case in which all diagrams are composed of a single coarse block, $\bar H^\alpha_n$.  
This occurs when there is a single index repeating an even number of times ($a_\mu=a$ for all $\mu$) or two unique indices, $a\neq b $, each repeating an odd number of times in the correlation function $\langle f_1\cdots f_m\rangle$. In the former case, we have shown that only the bare term remains. It follows that 
\begin{gather}
	\langle F_a \rangle = \langle F_a \rangle_0 + O(N^{-1/2}),
    \label{eq:ondiag}
\end{gather}
where we defined the local function $F_a$, which is simply a product of all the $f_\mu$, with index $a$:  
\begin{gather} 
F_a(t_1, \cdots, t_n) = \prod_{\mu: a_\mu =a} f_\mu(x_a(t_\mu),\hat x_a (t_\mu)),
\label{eq:foldeddefn}
\end{gather}
and $n$ is the number of times $a_\mu$ repeats (in this case $n=m$). Eq.\ \eqref{eq:ondiag} gives the leading order term in the expansion of $\langle f_1(x_a(t_1), \hat x_a(t_1)) f_2(x_a(t_2), \hat x_a(t_2)) \cdots f_m(x_a(t_m), \hat x_a(t_m))\rangle$. As one would expect, Eq.\ \eqref{eq:ondiag} states that the leading order behavior of moments involving a single neuron is given by the dynamical mean-field solution. 

We now consider the case in which there are exactly two distinct indices $a \neq b$ appearing in the correlation function $\langle f_1 \cdots f_m\rangle$. This requires summing over chain diagrams. Up to this point, we have been ignoring the time integral associated with each $J_{ij}$ gap. We must now treat these time integrals systematically. 

To do this, it is convenient to transform to Fourier space so that the convolutions between adjacent $J_{ij}$ gaps become products. To this end, define the response and $\phi$ average of the local $F_a$ functions appearing in the chain diagrams (i.e. the ``tufts" of Fig. \ref{fig:chains}) as 
\begin{gather}
    D_a(t_1-\tau, t_2 - \tau , \cdots, t_n - \tau) = \langle i \hat x_a(\tau) F_a(t_1, \cdots, t_n) \rangle_0, 
    \\ 
    P_a(t_1 - \tau, t_2 - \tau, \cdots, t_n - \tau ) = \langle \phi_a(\tau) F_a(t_1, \cdots, t_n) \rangle_0 , 
\end{gather}
where we used the stationarity of the mean-field background to write the moments in terms of time differences. Fourier transforming over the $t$ arguments turns the above into $e^{-i\tau \sum_\mu \omega^\mu} D_a(\bm \omega), e^{-i\tau \sum_\mu \omega^\mu} P_a(\bm \omega)$. The time integrals arising from the chains therefore reduce to products in frequency space evaluated at $\sum_\mu \omega^\mu.$  

We group the chains based on whether a factor of $\bm J \bm J^\top$ forms or not (Fig.\ \ref{fig:chains}; each row) and resum. The sum over each group of chains is in Fourier space 
\begin{align}
	&\langle F_a(\bm \omega_a) F_b(\bm \omega_b)\rangle 
    = 2\pi\delta\bigg(\sum_\mu \omega_a^\mu + \sum_\nu \omega_b^\nu\bigg) \bigg(
    \nonumber\\
    & h_1(\bm \omega_a, \bm \omega_b)
    \sum_{k\geq 1} (R^\phi_0(\bm \omega_a) \bm{J})^k 
    + h_2(\bm \omega_a, \bm \omega_b)
    ([R^\phi_0(\bm \omega_b) \bm{J}]^\top)^k
    \nonumber\\
	&
    + h_3(\bm \omega_a, \bm \omega_b)
    \sum_{k,l\geq 1} (R^\phi_0(\bm \omega_a) \bm J)^k 
    ([R^\phi_0(\bm \omega_b) \bm J]^\top)^l
    \bigg)^{\mathrm{no-repeats}}_{ab}
    \nonumber\\
    & +O(N^{-1}),
    \label{eq:fullsum}
\end{align}
where the scalar prefactors are 
\begin{align*}
    h_1(\bm \omega_a, \bm \omega_b)
    &= \frac{D_a(\bm\omega_a) P_b(\bm \omega_b)}{R^\phi_0(\bm \omega_a)},
    \\
    h_2(\bm \omega_a, \bm \omega_b)
    &= \frac{P_a(\bm\omega_a) D_b(\bm \omega_b)}{{{R}^\phi_0(\bm\omega_b)}},
    \\
    h_3(\bm \omega_a, \bm \omega_b)
    &= \frac{C^\phi_0(\bm \omega_a) D_a(\bm \omega_a) D_b(\bm \omega_b) }{R^\phi_0 (\bm \omega_a) R^\phi_0 (\bm \omega_b)},
\end{align*}
and we defined the bare two-point functions in Fourier space: $C^\phi_0(\bm \omega) = C^\phi_0(\sum_\mu \omega^\mu)$, and $R^\phi_0(\bm \omega) = \langle \phi'\rangle_0 /(1+i\sum_\mu\omega^\mu)$.

The ``no-repeats" label above denotes the index constraints imposed by the chain diagrams. That is, given a sequence of matrices $\bm{A}^{(1)},\cdots, \bm{A}^{(n)}$ we define the no-repeats product as 
\begin{widetext}
\begin{gather} 
\begin{aligned}
	(\bm A^{(1)} \cdots \bm A^{(n)})^{\mathrm{no-repeats}}_{ab} 
    &= \sum_{i_1 \neq i_2 \neq \cdots \neq i_{n-1} \neq (a,b)}  
    A^{(1)}_{ai_1}  A^{(2)}_{i_1 i_2} 
    \cdots  A^{(n)}_{i_{n-1} b}
\end{aligned}
    \label{eq:no-repeats-defn}
\end{gather}
\end{widetext}
with the sum over the $i$ variables, and we define $[AB]^{\mathrm{no-repeats}}_{ii} = 0$. Here, the constraint $i_1 \neq \cdots \neq i_{n-1} \neq (a,b)$ means that we sum over terms in which all of the $i$ indices are distinct from each other and from $a, b.$

We must now resum a geometric series subject to the no-repeats constraint. To motivate the analysis, note that the large-$N$ arguments that led us to the no-repeats constraint in the first place were largely combinatorial: they mostly relied on repeated index patterns and properties of i.i.d.\ random matrices, rather than on the details of the fields $\hat x$ and $\phi$. We therefore reapply the same logic directly to matrix products of the form $\bm J^k (\bm J^\top)^l$. 

To resum Eq.\ \eqref{eq:fullsum} we prove the following no-repeats lemma, valid for i.i.d.\ $J_{ij}$ obeying the moment assumption (Eq.\ \ref{eq:moment}). Let $p,q$ be integers with $p, q\geq 1$, then 
\begin{gather}
    \Jm^{p} (\Jm^\top)^{q} 
    = \big(\Jm^{p} (\Jm^\top)^{q}\big)^{\mathrm{no-repeats}}  
     + \sigma^2 \Jm^{p- 1 }(\Jm^\top)^{q-1}
    \\ + O(N^{-1/2} \bm I + N^{-1}\bm 1), \nonumber 
    \label{eq:J-lemma}
\end{gather}
and
\begin{gather}
    \bm J^k = (\bm J^k)^\mathrm{no-repeats} + O(N^{-1/2}\bm I + N^{-1}\bm 1), 
    \\ (\bm J^\top)^k = ( [ \bm J^\top]^k)^\mathrm{no-repeats} + O(N^{-1/2} \bm I + N^{-1} \bm 1).
\end{gather}
To show this, we drop the fields $\hat x, \phi$ from the definition of the diagrammatic symbols.  That is, we keep only the $\bm J$ in the definitions of the gaps in Fig.\ \ref{fig:defn}. The matrix product $\bm J^p (\bm J^\top)^q $ can now be represented as a sum over diagrams. In this notation, each diagram $D$ stands for the following sum: 
\begin{gather}
    D = \sum_{\mathbf{i} \in \mathcal A} J_{i_1 i_2} \cdots J_{i_{2k-1} i_{2k}},
\end{gather}
so that the matrix product follows from summing over equivalence classes of diagrams with a prescribed number and orientation of gaps. (Note that there are no symmetry factors here, as we are interested in a fixed, labeled ordering of the $J_{ij}$ gaps.) The same arguments used in Sec.\ \ref{sec:vardiags} now apply, leading to a sum over rainbows and chains. Expanding the innermost rainbow loop as $ \sigma^2 \bm I + O(N^{-1/2}) $ as in Sec.\ \ref{sec:cancellation} establishes the lemma.\footnote{Note that had we not defined $[AB]^{\mathrm{no-repeats}}_{aa} = 0$, we would be double counting the diagonals of $\bm J \bm J^\top$ products.}

Applying the lemma to Eq.\ \eqref{eq:fullsum} and imposing the constraint $\sum_\mu \omega^\mu_a + \sum_\nu \omega^\nu_b = 0$ generated by the delta function gives for $a \neq b$
\begin{align}
    \langle F_a&(\bm \omega_a) F_b(\bm\omega_b)\rangle 
    = \widetilde h_1(\bm \omega_a, \bm \omega_b)
    (\bm I - R^\phi_0(\bm\omega_a) \bm J)^{-1}_{ab}  
    \nonumber \\
    & + \widetilde h_2(\bm \omega_a, \bm \omega_b)
    (\bm I - R^\phi_0(\bm\omega_b) \bm J)^{-\top}_{ab}  
    \nonumber \\
    & + \widetilde h_3(\bm \omega_a, \bm \omega_b)
    [(\bm I - R^\phi_0(\bm\omega_a) \bm J)^{-1} 
    (\bm I - R^\phi_0(\bm\omega_b) \bm J)^{-\top }]_{ab}
    \nonumber \\
    & 
    + O(N^{-1}), 
  \label{eq:offdiagtwopoint}
\end{align}
where the scalar prefactors are now
\begin{align*}
    \widetilde h_1(\bm \omega_a, \bm \omega_b)
    &= \frac{D_a(\bm\omega_a) P_b(\bm \omega_b)}{R^\phi_0(\bm\omega_a)} 
    - \frac{C^\phi_0(\bm\omega_a) D_a(\bm\omega_a) D_b(\bm\omega_b) }
    {R^\phi_0(\bm\omega_a) R^\phi_0 (\bm \omega_b)},
    \\
    \widetilde h_2(\bm \omega_a, \bm \omega_b)
    &= \frac{  P_a(\bm\omega_a) D_b(\bm\omega_b) }{ R^\phi_0(\bm\omega_b)} 
    - \frac{C^\phi_0(\bm\omega_a) D_a(\bm\omega_a) D_b(\bm\omega_b) }
    {R^\phi_0 (\bm\omega_a) R^\phi_0(\bm\omega_b)},
    \\
    \widetilde h_3(\bm \omega_a, \bm \omega_b)
    &= \frac{D_a(\bm\omega_a) D_b(\bm\omega_b) }{R^\phi_0(\bm\omega_a) R^\phi_0(\bm\omega_b)}
    \\
    & \qquad C^\phi_0(\bm\omega_a) ( 1 - \sigma^2 R^\phi_0(\bm\omega_a) R^\phi_0(\bm\omega_b)).
\end{align*}
Note that we have assumed that the geometric series does not diverge, $\sigma |\langle \phi ' \rangle| < 1$; see Ref.\ \cite{shen2025covariance}. This formula, together with Eq.\ \eqref{eq:ondiag}, describes arbitrary two-point functions in the network. 

We now prove the result of Shen and Hu \cite{shen2025covariance}. Setting $F_a(t) = \phi(x_a(t))$ or $F_a(t) = x_a(t)$ and the same for $F_b$, Eqs.\ \eqref{eq:offdiagtwopoint} and $\eqref{eq:ondiag}$  can be written in matrix form in terms of a single frequency argument $\omega$ as: 
\begin{align}
    \bm C^\phi(\omega) 
    &= h_\phi(\omega) 
    (\bm I - R^\phi_0 \bm J )^{-1}  
    (\bm I - R^\phi_0 \bm J )^{-\dagger} 
    \label{eq:shen-hu}
    \\ 
    &\quad + O(N^{-1/2} \bm I + N^{-1} \bm 1 ), 
    \nonumber 
    \\
    \bm C^x(\omega) 
    &= h_x(\omega) \bm J (\bm I - R^\phi_0 \bm J)^{-1}
    (\bm I - R^\phi_0 \bm J)^{-\dagger} \bm J^\top 
    \label{eq:shen-hu2}
    \\ 
    & \quad   + \frac{G(\omega)}{1+\omega^2}  
    (\bm I - R^\phi_0 \bm J)^{-1}
    (\bm I - R^\phi_0 \bm J)^{-\dagger}
    \nonumber\\
    &\quad
      + O(N^{-1/2} \bm I + N^{-1} \bm 1 ).\nonumber
\end{align}
where
\begin{align*}
    h_\phi(\omega) &= C^\phi_0(1-\sigma^2 |R^\phi_0|^2),
    \\
    h_x(\omega) &= \frac{C^\phi_0 - \langle \phi'\rangle_0^2 C^x_0}{1 + \omega^2}.
\end{align*}
Here we introduced the shorthand $R^\phi_0 = \langle \phi'\rangle_0/(1+i\omega), R^x_0 =1/(1+i\omega)$. The averages $\langle \cdot \rangle_0$ and scalar correlation functions $C^\phi_0, C^x_0$ are averages of the two-point functions against the single-site mean-field theory in frequency space, Eq.\ \eqref{eq:mft-one}. In deriving the equation for $\bm C^x$ we have used the mean-field relation $C^x_0 = [\sigma^2 C^\phi_0 + G(\omega)]/(1+\omega^2)  $ together with the resolvent equation, $\Jm (\eye -\alpha \Jm)^{-1} = \alpha^{-1}[(\eye - \alpha \Jm)^{-1} - \eye]$, valid for arbitrary $\bm J$.

Finally, setting $F_a(t) = \phi(x_a(t))$ or $F_a(t) = x_a(t)$ and $F_b(t) = i\hat x_b(t) $ and using the fact that $ D_b(\omega_b) = - \langle \hat x^2\rangle_0 = 0$ gives the equations for the linear response functions: 
\begin{gather}
    \bm R^\phi(\omega) = R^\phi_0 (\bm I - R^\phi_0 \bm J)^{-1} + O(N^{-1/2} \bm I + N^{-1} \bm 1), 
    \\ 
    \bm R^x(\omega) = R^x_0 (\bm I - R^\phi_0 \bm J)^{-1}  + O(N^{-1/2} \bm I + N^{-1} \bm 1).
\end{gather}
This concludes our discussion of the two-point functions. 

We now turn to arbitrary $m$-point correlation functions. This requires summing over every way to assign the wavy lines into maximally disconnected blocks. In every such diagram, each index $a_\mu$ occurring an even number of times forms a single  coarse block $\bar H^\alpha$, while the indices $a_\mu$ repeating an odd number of times must be paired with one another into mixed blocks. We sum over every way of forming such mixed blocks. Defining $E$ to be the set of indices occurring an even number of times and $W$ to be the set of indices occurring an odd number of times, it follows that 
\begin{align}
	\langle f_1 \cdots f_m \rangle
	&= \prod_{e \in E} \langle F_e \rangle 
	\sum_{p \in P_2(W)} 
    \prod_{(\mu, \nu)\in p} 
    \langle F_\mu F_\nu\rangle 
    \nonumber \\
    &\quad + O(N^{-(|W|/4 + 1/2)})
    \label{eq:mpoint}
\end{align}
where $|W|$ denotes the number of indices in $W$, $P_2(W)$ is the set of all partitions of $W$ into pairs, and $F$ is defined as in Eq.\ \eqref{eq:foldeddefn}. The error in the residual is suppressed by a factor of $N^{-1/2}$ relative to the rest of the average. Combining the above with Eqs.\ \eqref{eq:ondiag} and \eqref{eq:offdiagtwopoint}, gives a closed solution to all intensive-order moments and their response functions, valid to the first non-trivial order in $N^{-1/2}$. The special case of all $a_\mu$ distinct then gives the simpler formula, matching Eq.\ \eqref{eq:simplemoment} of the introduction:
\begin{gather} 
\langle f_1\cdots f_m\rangle= \sum_{p \in P_2(1, \cdots, m)} \prod_{(\mu, \nu) \in p} \langle f_\mu f_\nu\rangle + O(N^{-m/4-1/2}).
\end{gather} 
This concludes the proof.

\section{Conclusion}

We have given a closed-form solution to arbitrary, intensive-order moments and response functions in a nonlinear random recurrent neural network, valid to leading nontrivial order in $1/\sqrt{N}$. While traditional dynamical mean-field theory approaches to random networks provide self-averaging statistics of the network activity, our results hold for a fixed instance of the quenched synaptic variables $J_{ij}$. 

In agreement with these results, Clark has developed an alternative cavity derivation of the two-point functions, Eqs.\ \eqref{eq:introshenhu} and \eqref{eq:responseintro} and subsequently used related diagrammatic arguments to obtain results on the higher cumulants; see Ref.\ \cite{clark2026linear}. We note in addition that a recent work, Ref.\ \cite{mato2025statistics}, derived a related Wick formula for intensive-order Fourier transformed moments at zero frequency $\omega=0$, after averaging over the $J_{ij}.$ Our results complement these findings by proving a general formula for intensive-order correlation and response functions, beyond the two-point functions, at a fixed instance of the quenched disorder $J_{ij}$.

Our derivation involves perturbatively expanding around the mean-field solution of the system, Eq.\ \eqref{eq:system}. This approach follows that of Thouless, Anderson, and Palmer (TAP) who first obtained fixed-disorder solutions of the Sherrington-Kirkpatrick spin-glass model by expanding around mean-field theory \cite{thouless1977solution}. Dynamical mean-field theory has been widely used to study the dynamics of large disordered systems. Notable examples include the Lotka-Volterra model of ecosystems \cite{roy2019numerical}, the Sachdev-Ye-Kitaev model \cite{PhysRevLett.70.3339, RevModPhys.94.035004}, and various spin glasses \cite{kirkpatrick1987dynamics, cugliandolo1993analytical, charbonneau2023spin}. A particularly relevant class of models for neuroscience are random recurrent neural networks with excitatory-inhibitory constraints on the synapses \cite{KadmonSompolinsky2015}. It is possible that the diagrammatic notation developed here can be used to analyze the behavior of these models at fixed instances of the quenched disorder.

\begin{acknowledgments}

I am very grateful to L.F.\ Abbott, Ashok Litwin-Kumar, Sebastian Mizera, David Clark, and Michael Mendelson for helpful discussions and comments on an earlier version of this manuscript. I also thank members of the Aronov lab for feedback on this work. 

This work was supported by the Howard Hughes Medical Institute, the Gatsby Foundation, and the National Institutes of Health grant T32MH126036. 

\end{acknowledgments}

\appendix

\section{Ordering lemma}

\label{sec:ordlemma}
Consider a nonzero diagram of an arbitrary order appearing in the expansion of the $m$th moment: 
\begin{widetext}
\begin{gather} 
\begin{aligned}
 \frac{i^k\symfactor}{2^l k! l!}
 \big\langle f_1 \cdots f_m 
 \sum_{\mathbf{i,j} \in \mathcal A}
 \hat x_{i_1} J_{i_1 i_2} \phi_{i_2}\cdots 
 \hat x_{i_{2k-1}} J_{i_{2k-1}i_{2k}} \phi_{i_{2k}} 
 \hat x_{j_1} \Delta \hat x_{j_1} \cdots 
 \hat x_{j_l}\Delta \hat x_{j_l} 
 \big\rangle_0 ,
\end{aligned}
\end{gather} 
\end{widetext}
where $ \mathcal A $ specifies index equality constraints, and we assume $f_\mu$ is one of $f_\mu(x,\hat x) = g_\mu(x), i\hat x$ with $g_\mu$ odd. We will prove that there exists an ordered partition of the $ J $ variables into $m/2$ blocks such that, within each block, neighboring $ J $ variables share an index, and the $ J $ variables on the ends share at least one index with one of the $a_1, \cdots, a_m$. This means that one can draw the diagram as a set of blocks such that within each block, neighboring $J_{ij}$ gaps have connected legs and wavy lines flank each block.

The proof of this relies on the fact that we need only consider diagrams in which every gap or bubble is connected by a sequence of edges to a wavy line. For example, consider the following diagrams 
\begin{center}
    \includegraphics[width=\linewidth]{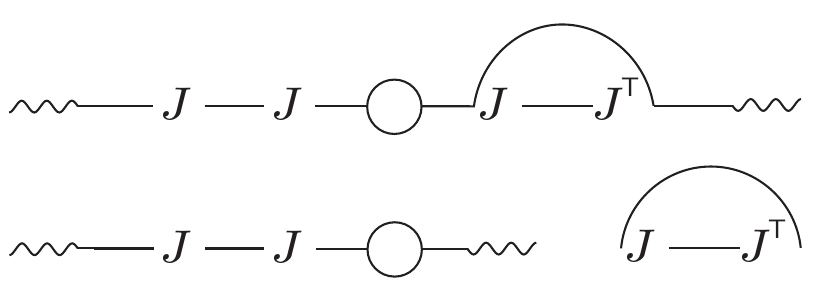}
\end{center}
\noindent In the top diagram, all $J_{ij}$ gaps and $\Delta$ bubbles are connected by a sequence of edges to wavy lines, while in the bottom diagram, there are no edges connecting the last two $J_{ij}$ gaps to the wavy lines. We will show that the sum over diagrams of the latter type is zero. 

If a diagram contains $J_{ij}$ gaps or $\Delta$ bubbles that are not connected by a sequence of edges to wavy lines, the corresponding sum can be factorized into an average involving the $f_\mu$ variables and an average that is independent of the $f_\mu$:
\begin{align} 
D 
&= \frac{\symfactor}{k!l!} 
\sum_{(\mathbf i, \mathbf j)\in \mathcal A^0} 
\langle f_1 \cdots f_m i\hat x_{i_1} J_{i_1 i_2} \phi_{i_2} \cdots  \frac 1 2 \hat x_{j_n} \Delta \hat x_{j_n}\rangle_0
\nonumber \\
&\quad
\sum_{(\mathbf i, \mathbf j) \in \mathcal A^1} 
\langle i \hat x_{i_1} J_{i_1 i_2} \phi_{i_2} \cdots \frac 1 2 \hat x_{j_{n'}}\Delta \hat x_{j_{n'}} \rangle_0
\end{align}
where the index set $\mathcal A^1$ must have no overlapping indices with $\mathcal A^0$, and $n+n' = l$ above. To show that the sum over all such diagrams is zero, we consider a given order of the expansion $k,l$, and we fix the part of the diagram that is connected to the  indices $f_1, \cdots, f_m$. Summing over all such diagrams gives a sum proportional to
\begin{gather} 
\begin{aligned}
 &\sum_{(\mathbf i, \mathbf j) \in \mathcal A^0} 
 \langle f_1 \cdots f_m i\hat x_{i_1} J_{i_1 i_2} \phi_{i_2} \cdots \frac 1 2 \hat x_{j_n} \Delta \hat x_{j_n}
 \rangle_0 
 \\
 &\quad
 \bigg\langle 
 \bigg(\sum_{i_1 i_2 \not \in \mathcal A^0} i\hat x_{i_1} J_{i_1 i_2} \phi_{i_2}\bigg)^\alpha 
 \bigg(\sum_{j\not \in \mathcal A^0} \frac 1 2 \hat x_j \Delta \hat x_j \bigg)^\beta
 \bigg\rangle_0 ,
\end{aligned}
\end{gather}
for some integers $\alpha, \beta$. Using the normalization of the MSRDJ functional
\begin{widetext}
\begin{align} 
 &\bigg\langle \bigg(\sum_{i_1 i_2 \not\in \mathcal A^0} i\hat x_{i_1} J_{i_1 i_2} \phi_{i_2}\bigg)^\alpha \bigg(\sum_{j \not \in \mathcal A^0} \frac 1 2 \hat x_j \Delta \hat x_j \bigg)^\beta \bigg\rangle_0 
 \nonumber \\ 
 &=
 \frac{\partial^{\alpha + \beta}}{\partial \lambda^\alpha \partial \rho^\beta} 
 \int \bigg( \prod_i Dx_i D \hat x_i\bigg)
 \exp \bigg(
 S_0[\bm x, \bm {\hat x}] 
 + \lambda  \sum_{i_1 i_2 \not\in \mathcal A^0} i\hat x_{i_1} J_{i_1 i_2} \phi_{i_2}
 + \rho  \sum_{j \not \in \mathcal A^0} \frac 1 2 \hat x_j \Delta \hat x_j
 \bigg) \bigg |_{\lambda, \rho =0}
 \nonumber \\
 &=
 \frac{\partial^{\alpha + \beta}}{\partial \lambda^\alpha \partial \rho^\beta}1 = 0. 
\end{align} 
\end{widetext}
This establishes the claim. The above argument corresponds to the vanishing of vacuum graphs in a Feynman diagram expansion (see e.g.\ Ref.\ \cite{hertz2016path}). In what follows, we therefore only consider diagrams in which every $J_{ij}$ gap and $\Delta$ bubble is connected to a wavy line by a sequence of edges. 

We now show that for the remaining diagrams, the $J_{ij}$ gaps can be ordered with neighbors sharing an index. Given that $\langle \hat x_j^{k}\rangle_0 =0$, the indices associated with $\Delta$ bubbles must be connected to an $a_\mu$ index or the index from a $J_{ij}$ gap. We therefore consider the case $l=0$ without a loss of generality. 

To establish the existence of such an ordering, define a multigraph in which each distinct index in $\mathcal A$ corresponds to a vertex and each edge corresponds to a $J_{ij}$ gap whose indices are set equal to the corresponding vertices. Since we only consider diagrams in which the $J_{ij}$ gaps are connected by a sequence of edges to wavy lines, every vertex in the resulting graph is connected to an $a_\mu$ vertex. 

We now use this graph to construct the diagram. To do this, we first note that for the $ \langle \cdot\rangle_0 $ averages over the $\hat x, x$ variables to be nonzero, every moment must be even, owing to the symmetry of the action $S_0[x,\hat x]$ under $(x, \hat x) \mapsto (-x, -\hat x)$. As such, each of the vertices which is not a fixed $a_\mu$ index must have an even degree, since each edge brings either an $i\hat x$ factor or a $\phi$ to the average. On the other hand, for indices $a_\mu$ repeating an odd number of times, the vertex must have an odd degree, as an odd number of $f_\mu$ enter into the expectation. By the same line of reasoning, fixed $a_\mu$ indices repeating an even number of times must have an even degree. 

\begin{figure*}[t]
    \centering
    \includegraphics[width=0.85\textwidth]{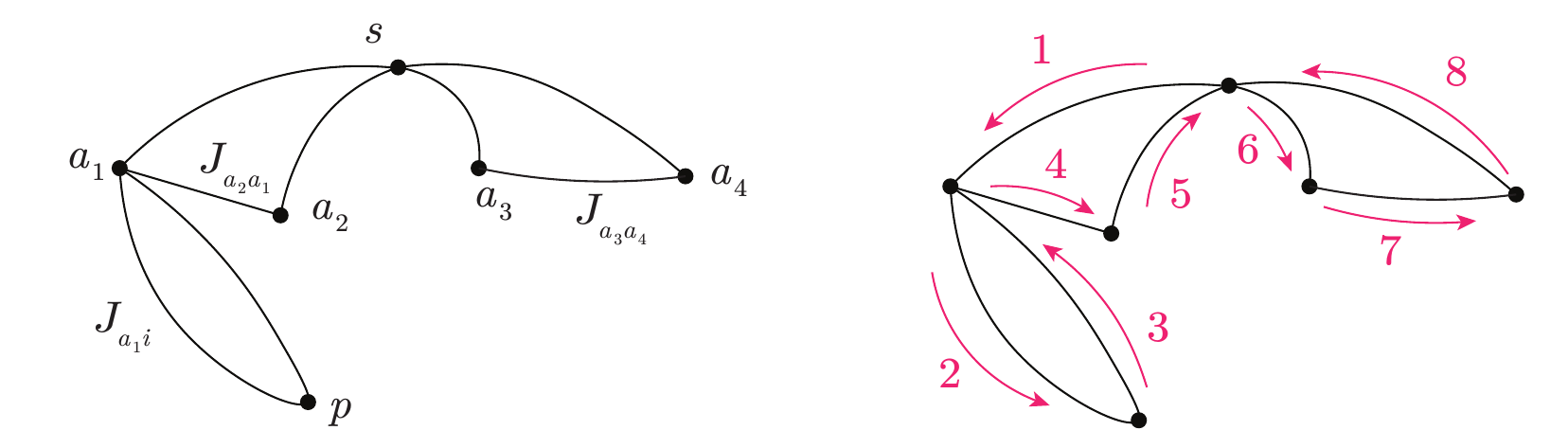}
    \label{fig:walk}
    \caption{Multigraph and walk for Eq.\ \eqref{eq:examplewalk}.}
\end{figure*}

Let us now attach an additional vertex, call it $ s $, to the multigraph, and connect it to each of the $ a_\mu $ vertices (see below for an example). For each repetition of $a_\mu$, we draw an edge between the $a_\mu$ vertex and $s.$ The result of this construction is a connected graph in which all of the vertices have an even degree after attaching $ s$. The Eulerian trail theorem then shows that there exists a walk through the graph visiting each edge---i.e. each factor of $i \hat x_{i_\alpha} J_{i_\alpha i_{\alpha+1}} \phi_{i_{\alpha+1}}$---exactly once. If we fix this walk to start and end at $ s $ we can see that it has the form 
\begin{gather} 
s\to W_1 \to s \to W_2\to \cdots  \to W_{m/2} \to s, 
\end{gather} 
where $ W_1, \cdots, W_{m/2} $ are disjoint walks through the graph that each start and end at vertices $ a_\mu $. These walks will be the $m/2$ blocks of the diagram. 

We now use the walks $ W_i$  to draw the diagram. For each $ W_i$, draw two wavy lines corresponding to the $a_\mu$ vertices visited at the start and end of the walk. Next, order the various $ J $ variables according to their appearance in the walk. Equivalently, for each walk $W_\alpha$ we form a set $H^\alpha_n$ consisting of index pairs. For the $n$th step of the walk, we set $H^\alpha_n=(i_\beta, i_{\beta+1})$ if the starting vertex corresponds to the left index of $J_{ij}$ and the ending vertex to the right, and we set $H^\alpha_n = (i_{\beta+1}, i_\beta)$ otherwise. This ensures neighbors share indices in an ordered fashion: $(H^\alpha_n)_2 = (H^\alpha_{n+1})_1$. Any additional connections between the $ J $ variables can then be drawn. We have established that any diagram can be drawn as a sequence of blocks with wavy lines flanking the ends of each block.

To illustrate this construction, consider the following term in the expansion of the four point function, $\langle \phi_{a_1} \cdots \phi_{a_4}\rangle$: 
\begin{gather}
\begin{aligned}
    \sum_{p  \neq (a_1, \cdots, a_4)}
    &\langle \phi_{a_1} \hat x_{a_1} \hat x_{a_1} \phi_{a_1} \rangle_0 
    J_{a_1 p}^2 
    \langle \phi_{p} \phi_{p}\rangle_0 
    J_{a_2 a_1} 
    \\
    &\langle \hat x_{a_2} \phi_{a_2} \rangle_0  
    \langle \phi_{a_3} \hat x_{a_3} \rangle_0 
    J_{a_3 a_4} 
    \langle \phi_{a_4} \phi_{a_4}\rangle_0, 
    \label{eq:examplewalk}
\end{aligned}
\end{gather}
with $a_1, \cdots, a_4$ all distinct. One obtains the  multigraph and walk shown in Fig.\ \ref{fig:walk}. Ordering the $J_{ij}$ following the prescription given by the walk and noting factors of $\bm J$ vs. $\bm J^\top$ following the above then gives the following diagram
\begin{center}
    \includegraphics[width=\linewidth]{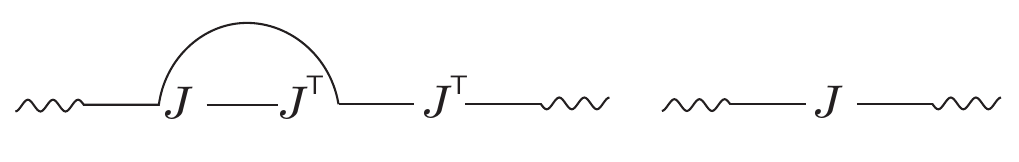}
\end{center}
\noindent This concludes the example. 

\section{Coarse-grained ordering}
\label{sec:coarse}

Consider a diagram in which an index repeats an arbitrary number of times: $a_\mu = a_\nu =\cdots = a_\gamma$. Any two distinct blocks $H^\alpha, H^\beta$ containing one of these repeated indices will then be connected. In this case it is possible that $Q_{\max}< m/2$. As such, maximally disconnected diagrams are those in which the repeated $ a_\mu  $ are grouped into the same blocks to the greatest extent possible.

We now show that for any maximally disconnected diagram, one can form a set of mutually disconnected ``coarse blocks" $\bar H^\alpha_n$ by stitching together $H^\alpha_n$ along shared indices. As we now show, the result of this procedure is a set of $Q_{\mathrm{max}}$ disconnected blocks, $\bar H^\alpha$, possessing the same ``neighboring vertices share an index" property as the original blocks $H^\alpha_n.$

To show this, consider a diagram containing $E$ ``same blocks," in which both wavy lines are associated with the same index, $a = a_\mu=a_\nu$, and $W$ ``mixed blocks," in which each wavy line corresponds to a distinct index. The claim follows trivially when all blocks are ``same blocks" so consider the setting in which $W\neq 0.$ Starting from a mixed block, $ H_0 $, stitch two ``same blocks" sharing the same external index to either side of $ H_0 $, if any exist. Since each fixed index appears in at most one mixed block for maximally disconnected diagrams, iterating this process eventually joins together all blocks connected to $ H_0 $. Repeating this procedure for all mixed blocks and doing the same for any remaining even blocks leaves us with a set of disconnected blocks $ \bar H^\alpha_n $ with $\alpha = 1,\cdots, Q_{\mathrm{max}} $. Crucially, the coarse blocks $\bar H^\alpha$ inherit the following property from $H^\alpha$: Given an element $ \bar H^\alpha_n $, the two neighbors $ \bar H_{n-1}^\alpha , \bar H^{\alpha}_{n+1} $ must share at least one index with $ \bar H^\alpha_n $.

\begin{figure*}
    \centering
    \includegraphics[width=0.85\textwidth]{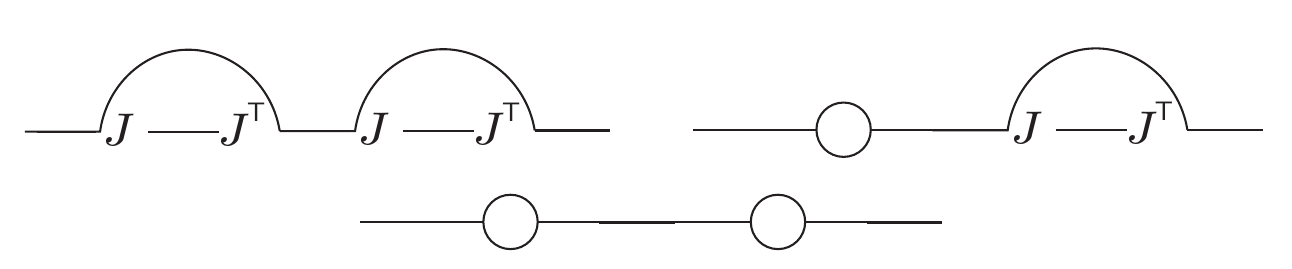}
    \caption{A rainbow and its decorations. Note that the diagram obtained by replacing only the right bare rainbow with a bubble is equivalent to the diagram in the upper right hand corner. Since we sum over equivalence classes of diagrams (Sec.\ \ref{sec:defns}), including both in the sum over decorations---i.e. counting the corresponding $c_\rho$ twice---would double count the corresponding term in the expansion of Eq.\ \eqref{eq:series} }
    \label{fig:dec}
\end{figure*}

\section{Combinatorial coefficients and cancellation}
\label{sec:cancel-comb}
In this section we prove the rainbow-bubble cancellation shown heuristically in the main text. To do this, we first identify the permutation symmetries that contribute to $\symsymbol$. We then establish the rainbow-bubble cancellation using an induction on the structure of the diagram. 

We begin by identifying the symmetries of a diagram. As described in the main text, given a diagram with $p$-paired $J_{ij}$ gaps, there are $2^p$ permutations between pairs that leave the pattern of connected indices unchanged. Additionally, note that exchanging two identical rainbows that are connected on their outermost edge similarly does not change the index connection pattern. For example the exchange between the two rainbows:
\begin{center}
    \includegraphics[width=\linewidth]{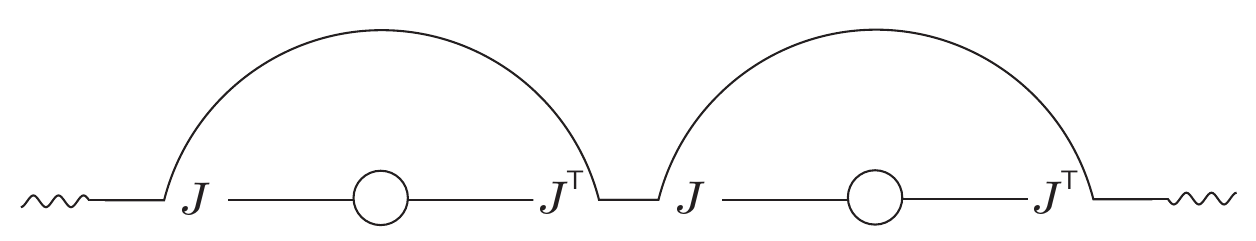}
\end{center}
We now rule out the possibility of additional symmetries. To do this, we begin by noting that any permutation that exchanges a $J_{ij}$ gap that is part of a chain and one which is part of a rainbow cannot be a symmetry, as the ``chain gap" cannot form a pair without violating the index constraints. 

Consider then an exchange between two $J_{ij}$ gaps, call them $\bar H^\alpha_n, \bar H^\alpha_m$, that are part of a chain. The two gaps must have at least one index distinct. When the gaps are swapped, this index in $\bar H^\alpha_{n}$ becomes connected with the corresponding neighbor of $\bar H^\alpha_m$. To avoid changing the pattern of connected indices, the neighbors must therefore be swapped with other gaps in the diagram. If the only suitable choice is a gap connected to a fixed index (wavy line) or a gap in a rainbow, then it is clear that the permutation is not a symmetry. Otherwise, if the issue can be fixed by a chain-chain exchange, then we may iterate this process until a wavy line constraint or a rainbow is reached. This rules out the possibility of permutations between gaps in a chain.

It remains to show that the only exchange between gaps that are each a part of a rainbow are the pairwise swaps and the exchanges between identical connected rainbows. 

We first note the following two constraints:
\begin{enumerate}
    \item Between two distinct connected rainbows the only shared indices are the boundary indices. 
    \item Between two distinct disconnected rainbows there are no shared indices. 
\end{enumerate}
From here we can proceed as before. Suppose two $J_{ij}$ gaps in rainbows $R$ and  $R'$ are swapped. If $R=R'$ and the two gaps are paired with each other, the swap is a symmetry, so suppose this is not the case. If the two gaps are connected boundary segments, then there is one index that is not shared between the two, otherwise both indices are distinct. The neighbor corresponding to the distinct index must be swapped, and the same considerations now apply. Proceeding in this way, one either hits a boundary segment or is forced to swap the neighbors to avoid a change in index connections. Iterating this process either leads to a full swap between identical rainbows, or it forces an exchange between a gap sitting inside of a rainbow and one sitting on its boundary. This is forbidden. Thus the only symmetries are pairwise exchanges and swaps between identical, connected rainbows. 

We now establish the cancellation between rainbows and bubbles. To begin, we recall that a \emph{rainbow} $R$ is a diagram built out of repeated nest and join operations (Sec.\ \ref{sec:leadingorddefn}). We define a \emph{decorated rainbow} as the diagram resulting from replacing some bare rainbows in $R$ by bubbles. We denote the pattern of decoration by $\rho$ and the set of all decorations by $\mathrm{Dec}(R)$ (Fig.\ \ref{fig:dec}).

Given a rainbow and its decoration, $(R,\rho)$, we define the \emph{coefficient} of the decorated rainbow to be $\alpha(R,\rho) =i^{p}/(2^q \symsymbol_{R, \rho})$, where $p$ is the number of gaps, $q$ the number of bubbles, and $\symsymbol_{R,\rho}$ the symmetry factor associated to the decorated rainbow. Finally, we let the sum over all coefficients be $W(R) = \sum_\rho \alpha(R,\rho)$. Since there are no additional symmetries between the rainbow and either chains or disconnected rainbows, it suffices to show that $W(R)=0$ for all $R$. 

This is done by induction on the number of join and nest operations---i.e., induction on the structure of the diagram. To start with, consider the base case. The bare rainbow clearly satisfies $W(R)=0$, since the bare rainbow has $\alpha = - 1/2$ and its decoration, the bubble, has $\alpha =1/2$. Let us then suppose that $W(R)=0$ for all rainbows constructed from $K$-many join and nest operations and consider a rainbow constructed from $K+1$ such steps. If the $K+1$ operation is a nesting under arcs, there are no new decorations to consider, and each $\alpha(R,\rho)$ simply is multiplied by an additional $i^2/2$ factor, so that $W(R)=0$ in this case.

Suppose then that the $K+1$ step is a join operation. The join operation in general can involve stitching together two rainbows that are themselves a joining together of more elementary rainbows, each constructed from $K'<K+1$ nest and join operations. Suppose that there are $L$-many identical rainbows joined together with the $i$th such rainbow repeating $n_i$ times. That is, the rainbow at this step has the form: 
\begin{gather} 
   \mathrm{Join}(\underbrace{R_1, \cdots, R_1}_{n_1 \mathrm{\ times}}, \cdots, \underbrace{R_L, \cdots, R_L}_{n_L \mathrm{\ times}})
\end{gather} 
with $R_i, R_j$ distinct for $i\neq j$, so that there are no symmetries between them. Consider some decoration of this rainbow. Each copy of an $R_i$ in the join receives some $\rho \in \mathrm{Dec}(R_i)$. Since we sum over equivalence classes of diagrams, the diagram is defined in terms of how many $R_i$ are decorated with each $\rho$ in $\mathrm{Dec}(R_i$)  (Fig.\ \ref{fig:dec}). That is, if we let $c_\rho$ denote the number of $R_i$ copies that receives that decoration (i.e. $\sum_\rho c_\rho = n_i$), we sum over each choice of $c$ vector only once to avoid double counting. Thus  we may write the total weight of the diagram as 
\begin{gather} 
\begin{aligned}
   \prod_i^l \sum_{c: \sum_\rho c_\rho =n_i} 
   \prod_\rho \frac {\alpha(R_i, \rho)^{c_\rho}} {c_\rho !}
   &= \prod_i  \frac 1 {n_i!}
   \bigg(\sum_{\rho} \alpha(R_i, \rho)\bigg)^{n_i} 
   \\
   &=  \prod_i \frac{W(R_i)^{n_i}}{n_i!} ,
\end{aligned}
\end{gather} 
where we used the multinomial theorem in the last step. But by the inductive hypothesis, $W(R_i)=0$ for all $R_i$. It follows that $W(R)=0$, concluding the proof.

\bibliographystyle{apsrev4-2}
\bibliography{references}

@article{hertz2016path,
  title={Path integral methods for the dynamics of stochastic and disordered systems},
  author={Hertz, John A and Roudi, Yasser and Sollich, Peter},
  journal={Journal of Physics A: Mathematical and Theoretical},
  volume={50},
  number={3},
  pages={033001},
  year={2016},
  publisher={IOP Publishing}
}

@article{bessis1980quantum,
  title={Quantum field theory techniques in graphical enumeration},
  author={Bessis, Daniel and Itzykson, Claude and Zuber, Jean-Bernard},
  journal={Advances in Applied Mathematics},
  volume={1},
  number={2},
  pages={109--157},
  year={1980},
  publisher={Academic Press}
}

@book{anderson2010introduction,
  title={An introduction to random matrices},
  author={Anderson, Greg W and Guionnet, Alice and Zeitouni, Ofer},
  number={118},
  year={2010},
  publisher={Cambridge university press}
}

@article{crisanti2018path,
  title={Path integral approach to random neural networks},
  author={Crisanti, A and Sompolinsky, H},
  journal={Physical Review E},
  volume={98},
  number={6},
  pages={062120},
  year={2018},
  publisher={APS}
}

@article{sompolinsky1988chaos,
  title={Chaos in random neural networks},
  author={Sompolinsky, Haim and Crisanti, Andrea and Sommers, Hans-Jurgen},
  journal={Physical review letters},
  volume={61},
  number={3},
  pages={259},
  year={1988},
  publisher={APS}
}

@article{shen2025covariance,
  title={Covariance spectrum in nonlinear recurrent neural networks},
  author={Shen, Xuanyu and Hu, Yu},
  journal={arXiv preprint arXiv:2508.05288},
  year={2025}
}

@article{mastrogiuseppe2018linking,
  title={Linking connectivity, dynamics, and computations in low-rank recurrent neural networks},
  author={Mastrogiuseppe, Francesca and Ostojic, Srdjan},
  journal={Neuron},
  volume={99},
  number={3},
  pages={609--623},
  year={2018},
  publisher={Elsevier}
}

@article{rajan2010stimulus,
  title={Stimulus-dependent suppression of chaos in recurrent neural networks},
  author={Rajan, Kanaka and Abbott, LF and Sompolinsky, Haim},
  journal={Physical Review E—Statistical, Nonlinear, and Soft Matter Physics},
  volume={82},
  number={1},
  pages={011903},
  year={2010},
  publisher={APS}
}

@article{KadmonSompolinsky2015,
    author        = {Kadmon, Jonathan and Sompolinsky, Haim},
    title         = {Transition to chaos in random neuronal networks},
    journal       = {Phys. Rev. X},
    volume        = {5},
    pages         = {041030},
    year          = {2015},
    doi           = {10.1103/PhysRevX.5.041030},
    eprint        = {1508.06486},
    archivePrefix = {arXiv},
    primaryClass  = {q-bio.NC}
}

@article{clark2023dimension,
  title={Dimension of activity in random neural networks},
  author={Clark, David G and Abbott, LF and Litwin-Kumar, Ashok},
  journal={Physical Review Letters},
  volume={131},
  number={11},
  pages={118401},
  year={2023},
  publisher={APS}
}

@article{roy2019numerical,
  title={Numerical implementation of dynamical mean field theory for disordered systems: Application to the Lotka--Volterra model of ecosystems},
  author={Roy, Felix and Biroli, Giulio and Bunin, Guy and Cammarota, Chiara},
  journal={Journal of Physics A: Mathematical and Theoretical},
  volume={52},
  number={48},
  pages={484001},
  year={2019},
  publisher={IOP Publishing}
}

@article{PhysRevLett.70.3339,
  title = {Gapless spin-fluid ground state in a random quantum Heisenberg magnet},
  author = {Sachdev, Subir and Ye, Jinwu},
  journal = {Phys. Rev. Lett.},
  volume = {70},
  issue = {21},
  pages = {3339--3342},
  numpages = {0},
  year = {1993},
  month = {May},
  publisher = {American Physical Society},
  doi = {10.1103/PhysRevLett.70.3339},
  url = {https://link.aps.org/doi/10.1103/PhysRevLett.70.3339}
}

@article{RevModPhys.94.035004,
  title = {Sachdev-Ye-Kitaev models and beyond: Window into non-Fermi liquids},
  author = {Chowdhury, Debanjan and Georges, Antoine and Parcollet, Olivier and Sachdev, Subir},
  journal = {Rev. Mod. Phys.},
  volume = {94},
  issue = {3},
  pages = {035004},
  numpages = {78},
  year = {2022},
  month = {Sep},
  publisher = {American Physical Society},
  doi = {10.1103/RevModPhys.94.035004},
  url = {https://link.aps.org/doi/10.1103/RevModPhys.94.035004}
}

@article{cugliandolo1993analytical,
  title={Analytical solution of the off-equilibrium dynamics of a long-range spin-glass model},
  author={Cugliandolo, Leticia F and Kurchan, Jorge},
  journal={Physical Review Letters},
  volume={71},
  number={1},
  pages={173},
  year={1993},
  publisher={APS}
}

@article{kirkpatrick1987dynamics,
  title={Dynamics of the structural glass transition and the p-spin—interaction spin-glass model},
  author={Kirkpatrick, Theodore R and Thirumalai, Devarajan},
  journal={Physical review letters},
  volume={58},
  number={20},
  pages={2091},
  year={1987},
  publisher={APS}
}

@book{charbonneau2023spin,
  title={Spin glass theory and far beyond: replica symmetry breaking after 40 years},
  author={Charbonneau, Patrick and Marinari, Enzo and Parisi, Giorgio and Ricci-Tersenghi, Federico and Sicuro, Gabriele and Zamponi, Francesco and Mezard, Marc},
  year={2023},
  publisher={World Scientific}
}

@article{mato2025statistics,
  title={Statistics of correlations in nonlinear recurrent neural networks},
  author={Mato, Germ{\'a}n and Rigatuso, Facundo and Torroba, Gonzalo},
  journal={arXiv preprint arXiv:2510.21742},
  year={2025}
}

@article{clark2026linear,
  title={Linear equivalence of nonlinear recurrent neural networks},
  author={David G. Clark},
  journal={arXiv preprint},
  year={2026}
}

@article{martin1973statistical,
  title={Statistical dynamics of classical systems},
  author={Martin, Paul Cecil and Siggia, Eric D and Rose, Harvey A},
  journal={Physical Review A},
  volume={8},
  number={1},
  pages={423},
  year={1973},
  publisher={APS}
}

@book{helias2020statistical,
  title={Statistical field theory for neural networks},
  author={Helias, Moritz and Dahmen, David},
  volume={970},
  year={2020},
  publisher={Springer}
}

@article{thouless1977solution,
  title={Solution of'solvable model of a spin glass'},
  author={Thouless, David J and Anderson, Philip W and Palmer, Robert G},
  journal={Philosophical Magazine},
  volume={35},
  number={3},
  pages={593--601},
  year={1977},
  publisher={Taylor \& Francis}
}

\end{document}